\documentclass[10pt,journal,compsoc]{IEEEtran}
\ifCLASSOPTIONcompsoc
  \usepackage[nocompress]{cite}
\else
  \usepackage{cite}
\fi
\ifCLASSINFOpdf
  \usepackage[pdftex]{graphicx}
  \DeclareGraphicsExtensions{.pdf,.jpeg,.png}
\else
\fi
\usepackage{amsmath}
\usepackage{amsthm,amssymb,bm}
\usepackage{array}
\usepackage{algorithm}
\usepackage{algorithmic}
\usepackage{color}

\ifCLASSOPTIONcompsoc
 \usepackage[caption=false,font=normalsize,labelfont=sf,textfont=sf]{subfig}
\else
 \usepackage[caption=false,font=footnotesize]{subfig}
\fi
\DeclareMathOperator*{\argmin}{argmin}

\newtheorem{theorem}{Theorem}
\newtheorem{lemma}{Lemma}

\begin{document}
%\title{Decompositional Quantum Graph Neural Network}

\title{Towards Quantum Graph Neural Networks:\\  An Ego-Graph Learning Approach}
\author{Xing~Ai, Luzhe~Sun,
Junchi~Yan,~\IEEEmembership{Senior Member,~IEEE,} Zhihong~Zhang$^{\ast}$, and~Edwin~R~Hancock,~\IEEEmembership{Fellow,~IEEE}
% <-this % stops a space
    \IEEEcompsocitemizethanks{
        \IEEEcompsocthanksitem Xing Ai is with School of Informatics, Xiamen University, Xiamen, Fujian, China. E-mail: 24320191152507@stu.xmu.edu.cn.
        \vspace{1mm}
        \IEEEcompsocthanksitem Luzhe Sun is with Computer Science Department in University of Chicago, Chicago, IL, US. E-mail: luzhesun@uchicago.edu.
        \vspace{1mm}
        \IEEEcompsocthanksitem J. Yan is with Department of Computer Science and Engineering, and MOE Key Lab of Artificial Intelligence, Shanghai Jiao Tong University, Shanghai 200240, China. E-mail: yanjunchi@sjtu.edu.cn. 
        \vspace{1mm}
        \IEEEcompsocthanksitem Corresponding author: Zhihong Zhang is with School of Informatics, Xiamen University, Xiamen, Fujian, China. E-mail: zhihong@xmu.edu.cn.
        \vspace{1mm}
        \IEEEcompsocthanksitem Edwin R. Hancock is with University of York, York, UK.  E-mail: edwin.hancock@york.ac.uk.}
        
% <-this % stops an unwanted space
}


\markboth{Journal of \LaTeX\ Class Files,~Vol.~14, No.~8, August~2015}%
{Shell \MakeLowercase{\textit{et al.}}: Decompositional Quantum Graph Neural Network}




\IEEEtitleabstractindextext{%
\begin{abstract}
Quantum machine learning is a fast-emerging field that aims to tackle machine learning using quantum algorithms and quantum computing. Due to the lack of physical qubits and an effective means to map real-world data from Euclidean space to Hilbert space, most of these methods focus on quantum analogies or process simulations rather than devising concrete architectures based on qubits. In this paper, we propose a novel hybrid quantum-classical algorithm for graph-structured data, which we refer to as the  Ego-graph based Quantum Graph Neural Network (egoQGNN). egoQGNN implements the GNN theoretical framework using the tensor product and unity matrix representation, which greatly reduces the number of model parameters required. When controlled by a classical computer, egoQGNN can accommodate arbitrarily sized graphs by processing ego-graphs from the input graph using a modestly-sized quantum device. The architecture is based on a novel mapping from real-world data to Hilbert space. This mapping maintains the distance relations present in the data and reduces information loss. Experimental results show that the proposed method outperforms competitive state-of-the-art models with only 1.68\% parameters compared to those models.
\end{abstract}

}

\maketitle

% is not selected <OR> if conference mode is selected - because compsoc
% conference papers position the abstract like regular (non-compsoc)
\IEEEdisplaynontitleabstractindextext

% \fi
\IEEEpeerreviewmaketitle

% \ifCLASSOPTIONcompsoc
% \IEEEraisesectionheading{\section{Introduction}\label{sec:introduction}}
% \else
\section{Introduction}
\label{sec:introduction}
% \fi

\IEEEPARstart{Q}{uantum} machine learning encapsulates a diverse variety of algorithms ranging from classical shallow learning techniques such as the Quantum Support Vector Machine (QSVM) \cite{Havlicek2019, Schuld2018} and the quantum decision tree classifier~\cite{QDT14} to the more recent quantum neural networks e.g. the Quantum Convolutional Neural Network (QCNN) \cite{Cong2018}, the Quantum Generative Adversarial Network (QGAN) \cite{Hueaav2761} and the Quantum Graph Neural Network \cite{qgnn}.

Distinct from the data processed by existing machine learning and deep learning techniques, the data used in quantum machine learning reside in high-dimensional Hilbert spaces represented in the form of quantum states. Maria et al. \cite{Maria2019} point out  that quantum machine learning algorithms overcome the problems associated with working in reduced dimensional space by using classical machine learning. Computation in high-dimensional Hilbert space can bring performance improvement. Seeking quantum counterparts of the Support Vector Machine (SVM) \cite{Vapnik1964}, Havlicek et al. \cite{Havlicek2019} propose a quantum binary classification algorithm similar to SVM (QSVM). A quantum circuit  is designed to map data from Euclidean space to Hilbert space. However, the input dimension cannot exceed three. Schuld et al. \cite{Schuld2018} propose a quantum-classical kernel method. A quantum computer estimates the inner products of the  data, while a classical computer calculates the estimation result and trains the algorithm. The Quantum Convolutional Neural Network (QCNN)  \cite{2018Modeling} simulates the structure of the  Convolutional Neural Network. For input sizes of $N$ qubits, the QCNN has only $O(\log{N})$ variational parameters. The Quantum Generative Adversarial Network (QGAN) of Hu et al.  \cite{Hueaav2761} has the potential for exponential acceleration relative to its classical counterpart. Hu et al. \cite{Hueaav2761} demonstrate that after multiple rounds of training, the generator of QGAN can generate the corresponding quantum states with 98.9\% fidelity. The generator is suitable for use on a medium-sized quantum computer in a noisy environment. 

Recent work has  aimed to combine quantum machine learning with graph representations~\cite{qgnn,QGCN,EQGC,QGRE,GraphQNTK}. Early work~\cite{qgnn, QGCN, EQGC} constructed quantum circuit based models  and showed these could successfully handle graph data. However, the input to these methods is the entire graph. The size of existing quantum devices is insufficient to handle this situation.  The above models can  thus only be applied to small-scale  or synthetic datasets. More recently, motivated by the  Graph Neural Tangent Kernel (GNTK)\cite{GNTK}, Tang and Yan propose a novel quantum kernel  for graph classification, referred  to as the  Graph Quantum Neural Tangent Kernel (GraphQNTK). This  is equivalent to  an infinite-width GNN with attention.

However, the above mentioned methods still suffer from two main limitations, namely  (i) they lack  a theoretical proof for being able to achieve graph isomorphism classification which we believe is a fundamental capability for expressive graph representation learning (in fact, we verify the discriminative ability of our model in our experiments) and  (ii) there is no existing  method for quantum graph learning that maps Euclidean data into  a quantum Hilbert space (and which we aim to address in our paper).

%{\color{red}{
%Despite the progress in developing the quantum RNN~\cite{QRNN20} and QCNN~\cite{Cong2018}, the quantum version of the Graph Neural Network (GNN)~\cite{scarselli2008graph} for real-world graph data has still not been well formulated or developed. In fact, the GNN has become an increasingly popular learning tool, especially for dealing with non-Euclidean data~\cite{non-Euclidean}. Emerging models and techniques include neighbor aggregation~\cite{xu2018powerful,GAT} which has found wide applications in social network analysis~\cite{NIPS2017_6703}, drug design~\cite{GCN_drag}, combinatorial optimization~\cite{NIPS2017_7214} and multi-agent learning~\cite{liu2020multi}. A quantum approach to the realization of GNN is therefore both conceptually attractive and imperative for dealing with large-scale graphs with high efficiency.}}

Graph-structured data or graph data are non-Euclidean  and consist of nodes and edges. It is widely used in knowledge representation \cite{ijcai2017-250}, social system analysis \cite{NIPS2017_6703}, \cite{kipf2016semi}, modelling higher-order interactions in physical systems \cite{pmlr-v80-sanchez-gonzalez18a}, \cite{NIPS2016_6418}, and combinatorial optimization \cite{NIPS2017_7214}. Recently, deep learning has achieved impressive results on a variety of tasks involving Euclidean data, and some studies have commenced generalizing deep learning to the graph domain. 

Graph Neural Networks (GNNs) \cite{Scarselli2009}, are recently proposed neural network structures for the processing of graph-structured data. The main idea underpinning GNNs is the neighborhood aggregation strategy. The strategy updates the representation of a node by recursively aggregating the representations of its neighbors. Xu et al.~\cite{xu2018powerful} prove that a GNN which satisfies certain conditions is as powerful as the Weisfeiler-Lehman (WL) test \cite{Weisfeiler1968} and can effectively distinguish the isomorphisms of graphs. Mathematically they prove GNNs can achieve the same effect as the WL test when three critical functions employed by  the GNN are injective. Different realizations of the GNN include but are not limited to Relational Graph Convolutional Networks (R-GCN) \cite{2018ModelingR-GCN},
%ERH define the meaning of edGNN
edge Graph Neural Networks (edGNN) \cite{jaume2019edgnn}, Random Walk Graph Neural Networks (RW-GNN) \cite{nikolentzos2020random} and Factorizable Graph Convolutional Networks (Factor GCN)~\cite{yang2020factorizable}. 

Recently, some studies have paid attention to introducing  an ego-graph into the  GNN to alleviate  the limitations of GNNs \cite{egognn} or to  provide insights into their performance\cite{Gophormer, EGI}. An ego-graph consists of a central node and all of its connected neighbors. To address the scalability issue while applying an  attention mechanism in  GNNs, Zhao et al. \cite{Gophormer}  adopt a transformer model for  ego-graphs rather than for  the entire graph. Moreover, Zhu et al. \cite{EGI} use ego-Graph information maximization to both analyze and provide theoretical guarantees on GNN transferability.

Motivated  by the above methods, this paper proposes a hybrid classical-quantum machine learning approach to Graph Neural Network embodiment. Compared to the existing quantum algorithm for graph-structured data, our method is scale-free and able to utilize the features of nodes. Additionally, we ground the GNN framework in the physical elements of quantum computing, i.e. qubits.

Our main contributions are summarized as follows:
\begin{itemize}
    \item [1)]
    A novel quantum-classical hybrid machine learning algorithm for graph-structured data is proposed, namely the Ego-graph based Quantum Graph Neural Network (egoQGNN). It utilizes the tensor product and unitary matrice representations to implement the theoretical framework of the GNN. Due to the fact that the  unitary matrix only requires the specification of a rotation angle, i.e. the variational parameter,  the egoQGNN achieves a similar performance but  with only 1.68\% of the  parameters  when compared with the GNN model with the least parameters, DGCNN~\cite{zhang2018end}.
    \item [2)]
    We design an ego-graph decompositional processing strategy to decompose a large graph into small ego-graphs which can be handled by existing small-sized quantum devices. By exploiting this strategy, the egoQGNN can handle larger-sized graph-structured data with a fixed-sized quantum device, via the efficient use of qubits. In the current situation where the number of physical qubits is limited, this is an important feature of our method.
    \item [3)]
    A trainable method for mapping data from a Euclidean space to a quantum Hilbert space is proposed. This can both maintain distance relations and also reduces information loss during the mapping.
\end{itemize}

The remainder of this paper is organized as follows. Section \uppercase\expandafter{\romannumeral2} reviews the existing GNNs and quantum machine learning algorithms. Section \uppercase\expandafter{\romannumeral3} introduces the fundamental concepts of quantum machine learning and GNNs. Section \uppercase\expandafter{\romannumeral4} introduces the trainable mapping method, the theoretical framework, and the resulting structure of the egoQGNN. Section \uppercase\expandafter{\romannumeral5} presents our experimental results obtained using the egoQGNN. Finally, Section \uppercase\expandafter{\romannumeral6} concludes the paper and offers directions for future investigation.

\section{Related Work}
In this section, we briefly review two related topics, namely 1) graph neural networks, especially those for graph level embedding; and 2) quantum machine learning. 

\subsection{Graph Neural Networks} 
The goal of graph neural networks is to map both node features and structural arrangement information to vectorized representations that can be used to embed graphs into lower-dimensional spaces or  manifolds. GNNs leverage node features and edge features to learn the representation of each node within a graph. By embedding the nodes of a graph and combining all the node embeddings within a graph, GNNs obtain the required graph representation. 

Since the seminal work in \cite{scarselli2008graph}, a diverse set of GNN models have been proposed.
%ERH you have already said this
For example, it has been proven~\cite{xu2018powerful} that a GNN that satisfies certain restrictive conditions can be as powerful as the Weisfeiler-Lehman (WL) test \cite{Weisfeiler1968} and can effectively distinguish the isomorphisms of graphs. Mathematically it has been proved that GNNs can achieve the same effect as the WL test when three critical functions used by the GNN  are injective. This analysis has been further extended to directed graphs by aggregating nodes and edges at the same time~\cite{jaume2019edgnn}. Spectrally-based GNNs \cite{SpectralNetworks,defferrard2016convolutional,rippel2015spectral} define spectral filter operations based on the Laplacian matrix of a graph. This is done by defining a series of filter coefficients based on the Laplacian eigenvalues and eigenvectors, and are thus computationally expensive. Fast and Deep Graph Neural Networks (FDGNN)~\cite{gallicchio2020fast} make it possible to combine the advantages of the deep architectural construction of GNNs with the extreme efficiency of  randomized neural networks, and in particular RC, methodologies. Random Walk Graph Neural Network (RWGNN)~\cite{nikolentzos2020random} generates graph representations by comparing a set of trainable {\it hidden graphs} against  input graphs using a variant of the $P$-step random walk kernel.

GNNs have achieved state-of-the-art results on different graph learning tasks, such as graph classification, link prediction and semi-supervised node classification. However, GNNs embed nodes into Euclidean space and cause significant distortion and structure. To overcome  this problem, several GNN models based on non-Euclidean geometry have been proposed \cite{NEURIPS2019_f8b932c7} \cite{chami2019hyperbolic} \cite{bachmann2019constant} \cite{liu2019hyperbolic}. These methods preserve the hierarchical structure of the graph and have improved performance compared to Euclidean GNNs. One common feature shared by   these methods and our model is  that of computing graph representation in a  non-Euclidean space.

\subsection{Quantum machine learning} 
Recently, there has been increasing interest in generalizing quantum computation to the machine learning domain. Several studies \cite{Vapnik1964,Schuld2018,Havlicek2019,2018Modeling,Hueaav2761} have been proven to be effective for classification. Unlike most classical machine learning methods, which process data in a Euclidean space, quantum machine learning methods map data to quantum states residing in a high dimensional complex-valued Hilbert space. Due to quantum entanglement, the dimensionality of the Hilbert space increases exponentially with the number of qubits. For example, the dimensionality of an $n$ qubit Hilbert space is $2^n$. This means that quantum states in this Hilbert space are $2^n$ dimensional complex vectors. In Section \uppercase\expandafter{\romannumeral3}, we will introduce the fundamentals of quantum machine learning in more detail.  Some observations that can be drawn from these methods are  (i) the inputs of these methods are quantum states defined in terms of qubits; (ii) the methods utilize quantum gates with learnable parameters to train the model and a classical computer is responsible for training, measuring and controlling quantum devices; (iii) the methods classify data by measuring the quantum state rather than applying softmax or similar functions. More details  given in Section \uppercase\expandafter{\romannumeral4}.

For graph structured data, some studies have attempted to utilize quantum machine learning to capture structural information and obtain graph or node representations in a  quantum Hilbert space. Verdon et al.~\cite{qgnn} were the first to propose a quantum computing based graph classification model, the so-called Quantum Graph Neural Network (QGNN). This captures   graph  structure using a  quadratic Hamiltonian and utilizing quantum circuits to extract  relevant graph structural information. With the aim of overcoming the  1-WL limitation of existing GNNs, P{\'e}ter et al. introduce  the Equivalent Quantum Graph Circuit (EQGC)~\cite{EQGC}. This captures the permutation-invariant topologies of  the input graphs. However, due to the number of required qubits scaling linearly with the number of nodes, EQGC can only handle small-scale synthetic datasets.  The Graph Quantum Neural Tangent Kernel (GraphQNTK) is the only known quantum algorithm that can handle realistically sized  graph data. Similar to  the Graph Neural Tangent Kernel (GNTK)~\cite{GNTK}, GraphQNTK is essentially a graph kernel and  is equivalent to an infinite-width GNN. 

The above quantum models have been successfully applied to graph-related tasks. However, most accept only low-dimensional data or pure quantum states as input. Two important problems limit the application of these methods to real-world large-scale data. Firstly, there is a lack of general methods to map data from a Euclidean space into a quantum Hilbert space. Secondly, due to the limited number of available qubits, real-world high-dimensional graph data can not be loaded into these quantum circuits. Besides, all  of the above methods lack  solid theoretical guarantees for the graph isomorphism classification problem. We also note that 
there are also related   quantum computing studies on  both node classification~\cite{QWNN,QSCNN,QGRE} and link prediction~\cite{HQGNN}. None of this work is grounded in a theoretical analysis or proof of graph isomorphism.
%To classify nodes in the attributed graph, Yan et al.\cite{QGRE} fed sampled node sequences by quantum random walk to quantum recurrent network to obtain node embeddings. 

In this paper, we propose a novel hybrid quantum-classical machine learning method referred to as the Ego-graph based Quantum Graph Neural Network (egoQGNN), which is a quantum computing based model for graph classification. By introducing the ego-network decompositional processing strategy (Sec.~\ref{Decompositional Processing Strategy}), the egoQGNN can be applied to real-world data and is not limited by the number of available qubits. The egoQGNN also implements the theoretical framework of GNNs using operations available in a quantum computer and utilizes a unitary weight matrix together with a  Hilbert-space tensor product. As a result, the egoQGNN empowers a quantum computer with the ability to classify graphs.

As shown in Table \ref{comparison}, as a hybrid quantum-classical algorithm, our model uses  a  hierarchical  architecture to capture information within $k$-hop neighbors of a node. Moreover, the proposed model is able to apply to graph isomorphism  tests since it provides a  theoretical analysis and proof of this capability  in Sec.~\ref{Theoretical Framework of egoQGNN}. With the proposed decomposition strategy, our model can handle real-world datasets.

\begin{table*}[tb!]
\centering
\caption{\centering{Comparison of the existing quantum graph representation learning methods, shows each model: is a hybrid quantum-classical algorithm or not, is able to apply to graph isomorphism or not, has hierarchy architecture or not, can handle real-world datasets or not}}
\begin{tabular}{ccccc}
\\\hline
Models& Hybrid& Graph Isomorphism& Hierarchy architecture& Real-World dataset\\ \hline
QGNN~\cite{qgnn}& No& No& Yes& No\\
QGCN~\cite{QGCN}& No& No& Yes& No\\
EQGC~\cite{EQGC}& Yes& No& No& No\\
QGCNN~\cite{QGCNN}& Yes& No& No& No\\
GraphQNTK~\cite{GraphQNTK}& Yes& No& No& Yes\\
egoQGNN(ours)& Yes& Yes& Yes& Yes\\ \hline
\end{tabular}
\label{comparison}
\end{table*}

\section{Preliminaries}
In this section, we briefly review the underlying concepts
which will be exploited in the development of a novel quantum neural network reported in this paper. Firstly, we introduce the fundamental concepts of quantum machine learning. Secondly, we review the fundamental theory of the GNN. In order to make our description clearer, we list the notation used in this paper in Table \ref{tab1}. 

\begin{table}[ht]
\centering
\caption{\centering{Notations and their descriptions}}
\begin{tabular}{|m{1.5cm}<{\centering}|m{5cm}|}
\hline
Symbol& Definition\\\hline
$h_v^t$& node $v$'s representation of $t$-th iteration\\\hline
$\mathcal{N}(v)$& a set of nodes adjacent to node $v$.\\\hline
$\sigma$& activation function\\\hline
$W^t$& weight matrix of $t$-th iteration\\\hline
$\vert\varphi\rangle, \langle\varphi\vert$& Quantum state and its complex conjugate transpose \\\hline
$U,U^\dagger$& arbitrary unitary matrix and corresponding adjoint\\\hline
$H$& Hadamard gate \\\hline
$X, Y, Z$& X gate, Y gate, Z gate \\\hline
$RX, RY, RZ$& rotation operators about the Pauli-X, Pauli-Y, Pauli-Z axes\\\hline
$\otimes$& tensor product\\\hline
\end{tabular}
\label{tab1}
\end{table}

\subsection{Fundamentals of Quantum Machine Learning}
In order to introduce our new quantum neural network more clearly, we will introduce some of  the fundamental elements of quantum machine learning including the quantum bit, quantum gate and quantum circuit. 

%quantum bit
\textbf{Quantum bit: }Quantum bits or qubits for short, are analogous to the classical binary bit which are the fundamental elements  of classical computation. The qubit is a mathematical construct with certain specific formal properties. Unlike the classical bit with binary states 0 or 1, the quantum state of a qubit is a 2-dimensional complex-valued vector formed by linear combinations of the basis vectors  $\vert0\rangle$ and $\vert1\rangle$:
\begin{equation}
    \vert0\rangle=(0,1)^\top \quad \vert1\rangle=(1,0)^\top
\end{equation}

Suppose the quantum state of a qubit is $\vert\varphi\rangle$:
\begin{equation}
    \vert\varphi\rangle=\alpha\vert0\rangle+\beta\vert1\rangle=(\alpha,\beta)^\top
\end{equation}

The numbers $\alpha$ and $\beta$ are complex numbers that satisfy the condition:
\begin{equation}
    \vert\alpha\vert^2+\vert\beta\vert^2=1\label{sumone}
\end{equation}

Due to the above condition in Eq\eqref{sumone}, the quantum state of a qubit is a point on the unit 3-dimensional sphere, referred to as the Bloch sphere. The Bloch sphere resides in a Hilbert space, and there are three basis vectors in this space, namely Pauli-X, Pauli-Y, and Pauli-Z. Any quantum state on the Bloch sphere makes angles with the three bases, as shown in Fig.~\ref{Bloch distance}.

There are an infinite number of points on the Bloch sphere. In other words, a qubit can represent an infinite number of quantum states, while a classical bit can only represent two states, i.e. 0 or 1. So, a qubit has a greater representational capacity than a classical bit. If a physical system, e.g. a quantum computer, has more than one qubit, for instance, $n$ qubits, then the quantum state of this physical system is the tensor product of all its constituent qubits or quantum states: 
\begin{equation}
    \vert\phi\rangle=\vert\varphi_1\rangle\otimes\vert\varphi_2\rangle\otimes\dots\otimes\vert\varphi_n\rangle
\end{equation}

The tensor product of $n$ qubits is a $2^n$-dimensional complex vector. So, the dimensionality of the  quantum state of a system increases exponentially with the number of qubits that constitute it.

%quantum gate
\textbf{Quantum gate: }
A classical computer has logic gates that change the states of bits, and these include the AND gate, OR gate, and NOT gate. For quantum computers, quantum gates are used to change the quantum states of the qubits. After the quantum  gate acts on the quantum state represented by a qubit, this quantum state is transformed, i.e. rotated, to give another vector. Quantum gates can be divided into single qubit gates and multiple qubit gates. 

Single qubit gates are applied to a single qubit state. Multiple qubit gates are applied to several qubit systems to transform their quantum states. Multiple qubit states can be obtained from the multiplication or tensor product of a single qubit gate. The transformation of states performed by the gate can be represented using a unitary matrix since the results of both a tensor product or a multiplication of quantum gates is a unitary transformation.
 
%quantum circuit
\textbf{Quantum circuits: }
A classical computer can be represented in terms of circuits containing connections and logic gates. Similarly, a quantum computer can be represented using quantum circuits containing connections and quantum gates. For a quantum computer, each connection represents a qubit, which is used to carry information, and the quantum gates perform manipulations to transform the quantum state. 
%In this paper, we use the form in Fig.~\ref{Fig.9} to represent quantum circuits, in which a connection represents a qubit and a block represents a quantum gate. The number of connections is the number of qubits contained in the system. The block superimposed on a single qubit represents a single qubit gate, while a block on several qubits represents a multiple qubit gate. 

\subsection{Graph Neural Networks on Graph Classification}
Graph Neural Networks (GNNs) are effective machine learning tools for structured data and rely on a neighborhood aggregation strategy. Specifically, for each node in a graph, the GNN recursively aggregates the current representation with those for its neighbors, thus giving a new representation for use at the next iteration. For graph classification, Xu et al.~\cite{xu2018powerful} demonstrate that three crucial functions of GNNs, namely AGGREGATE, COMBINE and READOUT, must each be injective multi-set functions, for example, a sum. Formally, the $k$-th interaction of a GNN can be expressed as:
\begin{equation}
    \begin{cases}
    h_{\mathcal{N}(v)}^{k}=AGG^{(k)}(\{h_\mu^{(k-1)},\forall\mu\in\mathcal{N}(v)\})\\
    \\
    h_{v}^{k}=\sigma(W^k\cdot COM^{(k)}(h_v^{k-1},h_{\mathcal{N}(v)}^{k}))
    \end{cases}
\end{equation}
where $h_{v}^{k}$ is the feature vector of node $v$ at the $k$-th iteration, and $\mathcal{N}(v)$ is the set of nodes adjacent to $v$. 
 
Here, the READOUT function aggregates the representations of all nodes from the final iteration to obtain the complete graph representation $h_G$:
\begin{equation}
    h_G=READOUT(\{h_v^K\vert v \in G\})
\end{equation}

The GNN model presented by Xu et al.~\cite{xu2018powerful} can be represented as follows:
\begin{equation}
    h_{v}^{k}=\sigma(h_v^{(k-1)}W_1^{(k-1)}+\Sigma_{\mu\in\mathcal{N}(v)}h_\mu^{(k-1)}W_2^{(k-1)})\label{Eq.3}
\end{equation}
where $W_1^{(k-1)}$ and $W_2^{(k-1)}$ are trainable weight matrices at the $k$-th iteration for node $v$ and its neighbors respectively.

\section{Methodology}
\label{method}
In this paper, we develop a novel quantum-classical hybrid machine learning algorithm for graph-structured data. The idea is to develop a quantum GNN by designing a quantum circuit and replacing the Euclidean weight matrices of the GNN with unitary matrices, i.e. quantum gates. In this way, we incorporate theoretical ideas from quantum machine learning into deep learning in the graph domain. 

In this section, we first introduce the theoretical framework underpinning our model and provide mathematical proof. Then, we introduce the quantum circuit to implement  the egoQGNN, the trainable mapping method and the sequential decompositional processing strategy. Finally, we introduce the structure and explain how the egoQGNN can be used to classify quantum states. We then summarize the processes underpinning the  egoQGNN.

\subsection{Theoretical Framework of egoQGNN}
\label{Theoretical Framework of egoQGNN}
To distinguish graph isomorphisms by quantum machine learning, we follow the study of Xu et al.~\cite{xu2018powerful}, and introduce the theoretical framework  for GNNs using quantum machine learning. What makes the  GNN so powerful is its injective aggregation strategy, which maps different nodes to different representational units. As a result, different graphs have different representations. 

The GNN model \cite{xu2018powerful} can be represented by:
\begin{equation}
    \mathbf{h}_{v}^{k}=\sigma\left(\mathbf{h}_v^{(k-1)}\mathbf{W}_1^{(k-1)}+\Sigma_{\mu\in\mathcal{N}(v)}\mathbf{h}_\mu^{(k-1)}\mathbf{W}_2^{(k-1)}\right)\label{GIN}
\end{equation}
where $\mathbf{h}_v^t$ is the representation of node $v$ at the $t$-th iteration and $\mathcal{N}(v)$ is the neighbor set of node $v$. The matrices $\mathbf{W}_1^{(k-1)}$ and $\mathbf{W}_2^{(k-1)}$ are respectively trainable weight matrices at the $(k-1)$-th layer for node $v$ and its neighbors;  $\sigma$ is the  network activation function.

The aim of this paper is to provide a route to implementing Eq.~\ref{GIN} on a quantum computer. For quantum computing, the state of a physical system composed of several qubits is obtained by a tensor product of the individual qubit quantum states, namely quantum entanglement. The tensor product is one of the fundamental constructs of quantum computing. Due to its  ability to enlarge the space exponentially, the tensor product can be used to map different nodes to different representations. The tensor product is analogous to the summation operation in classical GNNs. In other words, both the tensor product and the summation are injective functions. 

\begin{lemma}\label{lemma:1}
The tensor product is injective, for non-zero vectors with the same dimension. As a result, the tensor product maps them to different representations unless they are linearly dependent vectors.
\end{lemma}

All the proofs including that for this Lemma can be found in the Appendix. Since quantum states are complex vectors, according to Lemma 1, the tensor product maps different quantum states to different representations. Moreover, all quantum states are linearly independent. 

\begin{lemma}\label{lemma:2}
If two quantum states $\mathbf{A}$ and $\mathbf{B}$ are linearly dependent, and $\mathbf{B}=k\cdot \mathbf{A}$, then $k= \pm1$.
\end{lemma}

Due to the property of quantum states, Lemma 2 can be demonstrated easily using the proof in the Appendix. According to Lemma 1 and Lemma 2, the tensor product maps two quantum states to the same representation if and only if the two quantum states are identical.

So, for the egoQGNN, the tensor product replaces the summation of the GNN, and Eq.~\ref{GIN} becomes:
\begin{equation}
    \mathbf{h}^{t-1}(v)\to\vert\varphi_v\rangle^{t-1}, \quad \mathbf{h}^{t-1}(\mu)\to\vert\varphi_\mu\rangle^{t-1}\label{node quantum state}
\end{equation}
\begin{equation}
    \vert\varphi_{v}\rangle^{t}=\mathbf{U_1}\vert\varphi_{v}\rangle^{t-1}
    \bigotimes\left(\bigotimes_{\mu\in\mathcal{N}(v)}\mathbf{U_2}\vert\varphi_{\mu}\rangle^{t-1}\right) \label{node representation}
\end{equation}
\begin{equation}
    \mathbf{h}^{t}(v)=-\text{tr}(\mathbf{\rho}^{t} \log\mathbf{\rho}^{t}), \quad \mathbf{\rho}^{t}=\vert\varphi_{v}\rangle^{t}\langle\varphi_{v}\vert^{t} \label{next representation}
\end{equation}
where $\mathbf{U}_1$ and $\mathbf{U}_2$ are unitary matrices with trainable parameters, which correspond to the Euclidean weight matrices $\mathbf{W}_1$ and $\mathbf{W}_2$ in the GNN of Eq.~\ref{GIN}. Unlike the weight matrices of the  classical GNN which have many numerical parameters which must be  trained, the unitary matrices have only a single variational parameter that represents the  rotation angles of the quantum states. Details regarding  unitary matrices and their properties are given in the  Appendix.

Note  that Eq.~\ref{node representation} implements on a quantum computer and returns the result to a classical computer. We will discuss the implementation of Eq.~\ref{node representation} in the next section. Eq.~\ref{node quantum state} is achieved by the proposed mapping method. The representation of the next layer can be obtained by von Neumann entropy in Eq.~\ref{next representation}.

Both Eq.~\ref{GIN} and Eq.~\ref{node representation} are able to distinguish non-isomorphic graphs, i.e. perform graph classification. To demonstrate the effectiveness of the egoQGNN for the graph isomorphism problem, we need to prove that Eq.~\ref{node representation} will map nodes with different features or neighbors to different representations. 

\begin{lemma}\label{lemma:3}
For different nodes $v$ and $\mu$, the outputs of Eq.~\ref{node representation}: $\vert\varphi_v\rangle$ and $\vert\varphi_\mu\rangle$, meet  $\vert\varphi_v\rangle\neq\vert\varphi_\mu\rangle$.
\end{lemma}

A natural follow-up theorem is that the egoQGNN can distinguish graphs that are decided as non-isomorphic graphs by the GNNs based on Eq.~\ref{GIN}. 

\begin{theorem}\label{theorem:1}
The egoQGNN maps two non-isomorphic graphs $G_1$ and $G_2$ as decided via a  GNN by Eq.~\ref{node representation} to different embeddings.
\end{theorem}

%Again, we will give the proofs of the above lemma and theorem in the Appendix.

Compared with the GNN, the main differences provided by  the egoQGNN are the tensor product aggregation operator and the unitary matrix. The advantages of using the tensor product and unitary matrix are as follows:

1) The tensor product enlarges the node representation space exponentially. As a result, nodes with different features can be mapped to different representations. Although alternative  functions such as the dot product and matrix multiplication are both injective functions, their implementations on quantum devices are not as convenient as the tensor product. This is  because the tensor product is a fundamental facet of quantum computing.

2) Since the tensor product can enlarge the representation space exponentially, the egoQGNN has significantly fewer parameters than related deep learning models. A layer of the GNN requires a $d\times h$ weight matrix to transform a $d$-dimensional input into a $h$-dimensional output. Such a layer has $d\times h$ parameters that need to be trained. For  the egoQGNN, the entanglement of $max(\log_2d,\log_2h)$ qubits can generate  $max(d,h)$-dimensional quantum states. The egoQGNN has only $n\cdot max(\log_2d,\log_2h)$ variational parameters for training if $n$ quantum gates are applied to each qubit. This is because a unitary matrix has only one variational parameter (the rotation of the quantum states).

\begin{figure}[tb!]
    \centering
    \includegraphics[scale=0.2]{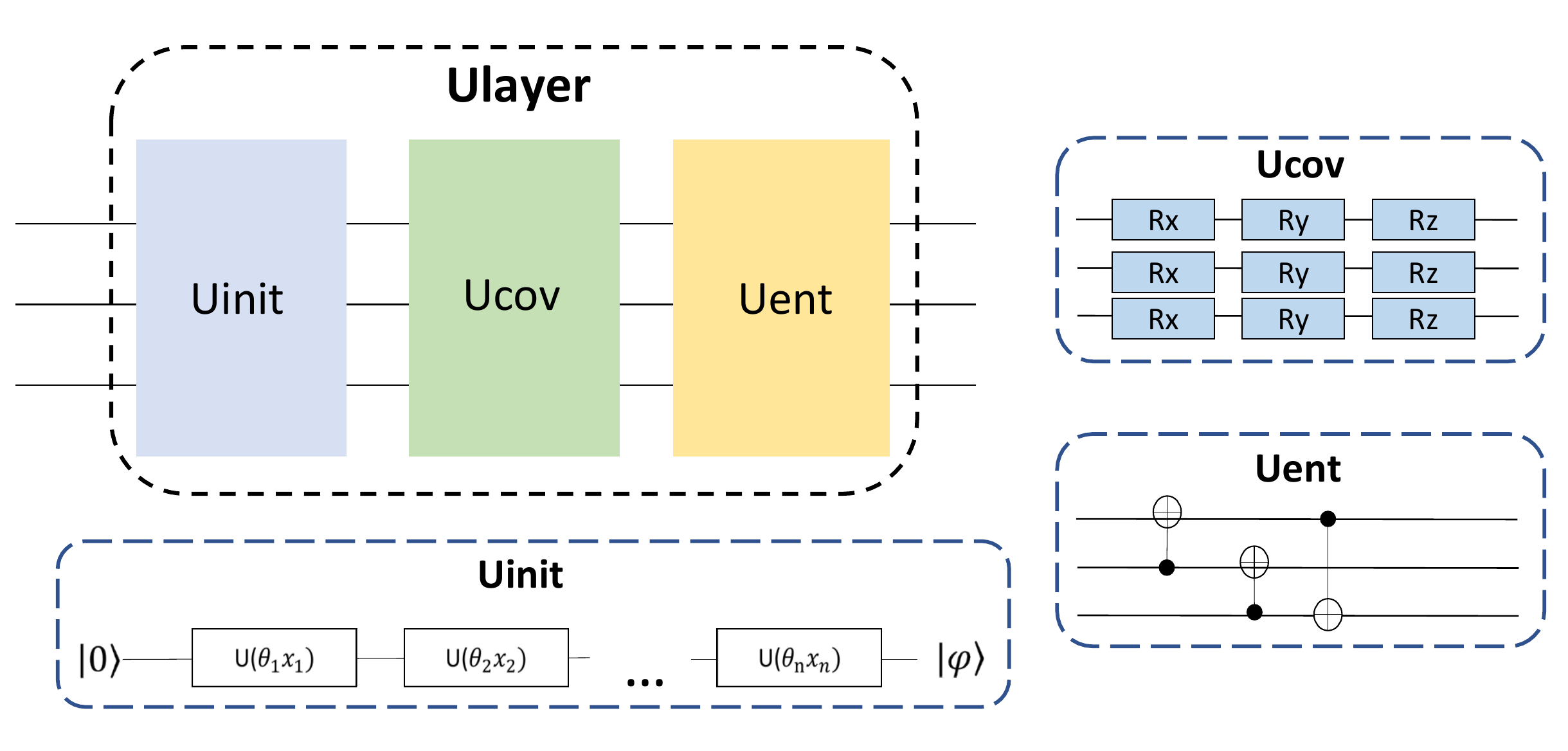}
    \vspace{-10pt}
    \caption{Ulayer with Uinit, Ucov, and Uent. Uinit contains $n$ quantum gates to map $n$-dimensional data $X$ into quantum states. In Ucov, RX, RY, RZ quantum gates are applied to each qubit. Uent utilizes CNOT gates to entangle all qubits. }
    \label{Ulayer}
\vspace{-10pt}
\end{figure}

\subsection{Quantum Circuit of egoQGNN}
\label{quantum circuit}
To implement Eq.~\ref{node representation} on a quantum device, we design a quantum circuit with a hierarchical structure similar to  that of the GNN. The quantum circuit we have designed contains $K$ Ulayers to capture the information within the $K$-hop neighbors of a node. A Ulayer includes three components: Uinit, Ucov and Uent. This is illustrated  in Fig.~\ref{Ulayer}. For $n$  input nodes, the initial states of the quantum circuit are $\vert0\rangle^{\otimes n}$. The output is the tensor product of the quantum states over all the  qubits.

We allocate a quantum circuit and the requisite qubits to the ego-graphs consisting of a node and its neighbors. After the required computations have been completed, we then free the quantum circuit and its qubits for future use. 

The Uinit component maps the data to quantum states, which corresponds to the quantum circuit of the trainable mapping method (described  later in this section). Specifically, it is responsible for mapping node features from Euclidean space to Hilbert space. The Ucov component, on the other hand, has three quantum gates on each qubit: $\mathbf{RX}$, $\mathbf{RY}$, $\mathbf{RZ}$. Each quantum gate accepts one trainable parameter. All neighbors of a node share parameters but do not include the node. For  an arbitrary qubit of the Ucov component, if the input quantum state is $\vert\varphi_{in}\rangle$, the output is $\vert\varphi_{out}\rangle$:
\begin{equation}
    \vert\varphi_{out}\rangle=\mathbf{U}_{cov}\vert\varphi_{in}\rangle=\mathbf{RX}(\theta_x)\mathbf{RY}(\theta_y)\mathbf{RZ}(\theta_z)\vert\varphi_{in}\rangle
    \label{inout}
\end{equation}

The $\mathbf{U}_{cov}$ corresponds to $\mathbf{U}_1$ or $\mathbf{U}_2$ in Eq.~\ref{node representation}. $\theta_x, \theta_y, \theta_z$ are variational parameters of quantum gates. The output of Ulayer is a tensor product over the quantum states of all nodes, corresponding to $\vert\varphi_v\rangle^{t}$ in Eq.~\ref{node representation}. Therefore, the Ulayer component implements Eq.~\ref{node representation}. In a manner similar to the GNN, the egoQGNN can aggregate features of the $k$-hop neighbors to a node by applying $K$ Ulayers.  Ucov is followed by a Uent component, which applies a CNOT gate to each pair of qubits to entangle their information. The parameters of the Uinit component have been trained and will not be updated during the training of Ulayer.

%The quantum gates of DQGNN are few, resulting in introducing less noise. 

\begin{figure}[tb!]
    \centering
    \includegraphics[scale=0.14]{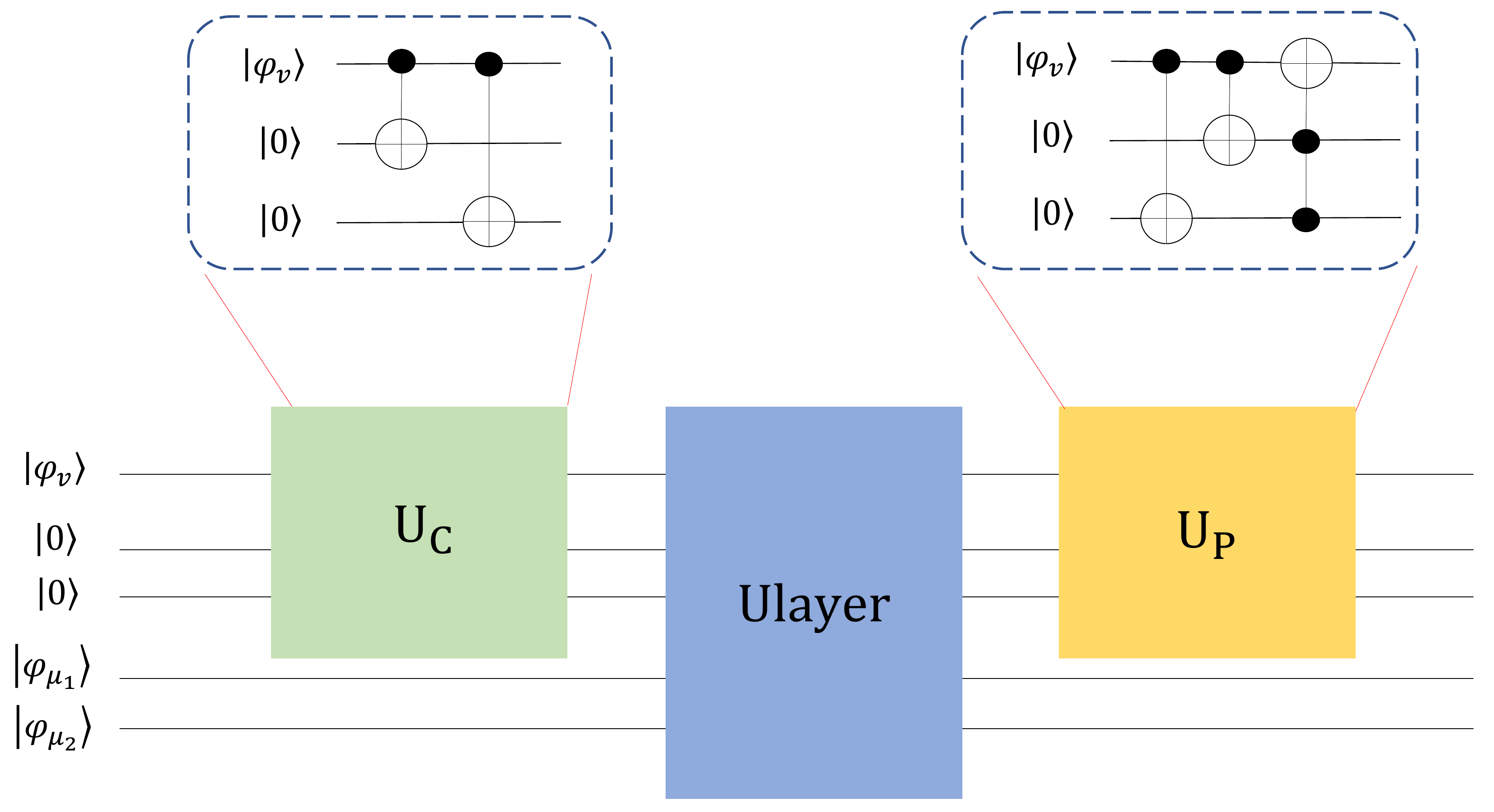}
    \vspace{-10pt}
    \caption{Following the three-bit error correction code proposed by \cite{raussendorf2012key}, the error correction of egoQGNN can be achieved by applying $U_c$ and $U_p$ modules after and before Ulayer respectively. }
    \label{QEC}
\vspace{-10pt}
\end{figure}

Considering the likely  effects of noise interference  on quantum devices, we apply the three-bit error correction code\cite{raussendorf2012key} to the above circuit. According to Eq.\eqref{node representation}, the goal of the egoQGNN circuit is to  aggregate the neighboring quantum states ($\vert\varphi_{\mu}\rangle$) into the quantum state of the node $v$ ($\vert\varphi_{v}\rangle$) to update the representation of the node $v$. To avoid the interference of noise on the representation of the node $v$, the components $U_C$ and $U_P$  are respectively applied before and after the application of the  Ulayer. Specifically, $U_C$ first copies the information from the target bit to the two auxiliary qubits via two CNOT gates. After applying the  Ulayer, $U_P$ applies three CNOT gates to assist the  recovery of  the state of  the target qubit $\vert\varphi_{v}\rangle$, as shown in Fig.~\ref{QEC}.

\subsection{Trainable Mapping Method}
\label{trainable mapping}
Existing quantum machine learning techniques  \cite{Havlicek2019,Hueaav2761,Rebentrost2014,Cong2018, 2018Quantum,SITU2020193}
are mainly applied to artificial data or small-scale real-world data. One reason for this restriction is the lack of an effective mapping from a variety of real-world data types to quantum states. To address this problem, we propose trainable mapping to maintain the distances between data and to reduce information loss resulting from the  mapping. 

\begin{figure}[t!]
    \centering
    \includegraphics[scale=0.14]{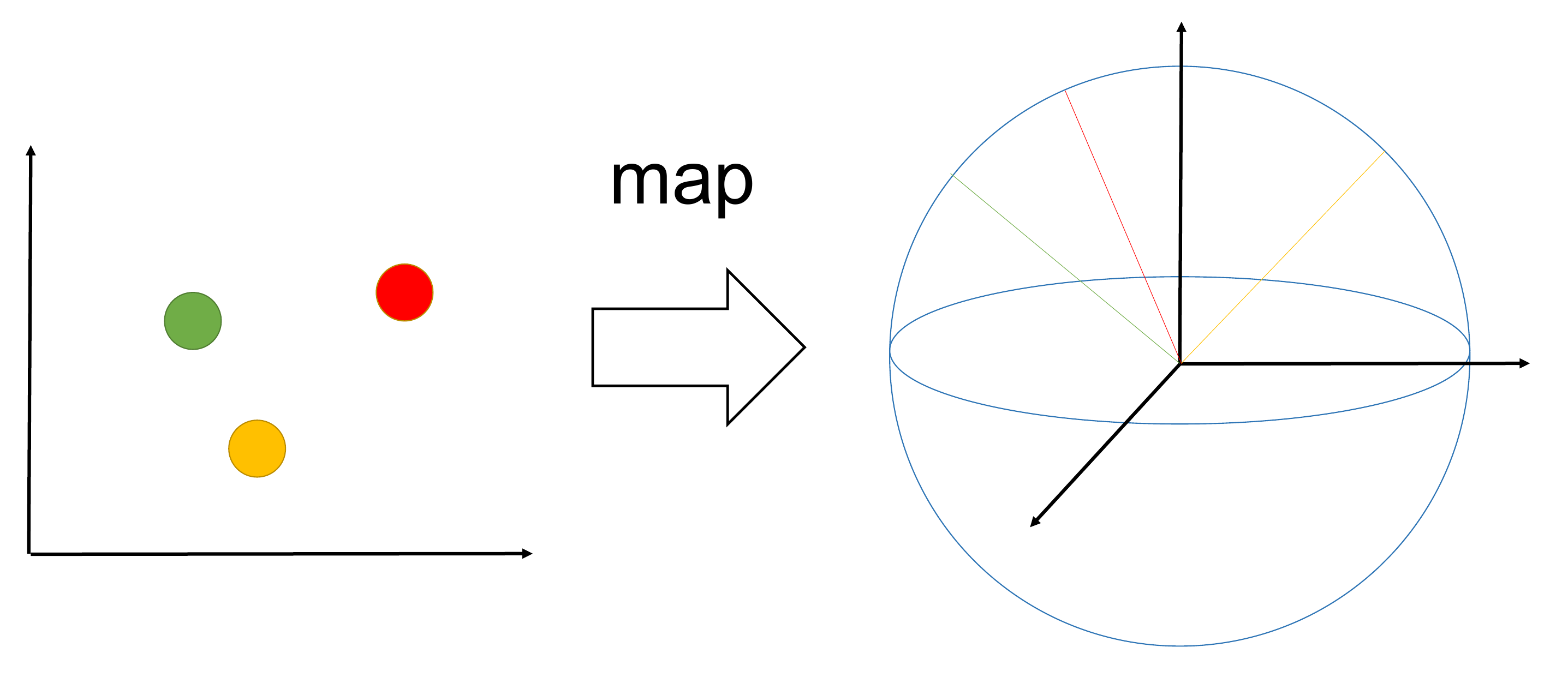}
    \vspace{-10pt}
    \caption{In Euclidean space, the distance between the green point and the orange one is smaller than that between the green and the red ones. After mapping the data points to Hilbert space, points close to each other in Euclidean space become more distant  quantum states (the green and orange lines), while the distance between the points which are farther away in Euclidean space becomes closer (the red and green lines).}
    \label{Bloch distance}
\vspace{-10pt}
\end{figure}

Quantum states are distributed on the surface of the Bloch sphere that resides in Hilbert space. Mapping data from a  Euclidean space to the Bloch sphere may incur a large distortion, as shown in Fig.~\ref{Bloch distance}.

The distance relations between the quantum states on the Bloch sphere are expected to be consistent with those between the data in Euclidean space. Any two sample points that are close to each other in Euclidean space should also be close after the mapping to Hilbert space. If this is not the case, the distance difference between the quantum states and the data will induce information loss and impair the subsequent analysis operations. Considering the above problem, we present a mapping method in which the loss is related to the difference in the distance relations between the Euclidean space and the Bloch sphere. The corresponding circuit implementation for this method is shown in the Uinit component of Fig.~\ref{Ulayer}. For $n$-dimensional data $\mathbf{x}=(x_1,x_2,\dots,x_n)$, the mapping circuit repeatedly uses a qubit with $n$ quantum gates to map each sample to a corresponding quantum state. Each quantum gate accepts the product of the $x_i$ and the trainable parameter is $\theta_i$. Obviously, the Euclidean distance is not suitable for distance measurement on the Bloch sphere, which is a unit sphere. We utilize the inverse cosine of the inner product of the quantum states for distance measure. The reason that we chose inverse cosine is the low computational complexity and natural consistency with the spherical representation. We believe other non-Euclidean distance measures are also available for our model, e.g. hyperbolic distances.

We construct a loss based on the correlation matrix $\mathbf{D}$ and $\mathbf{D}^{'}$ of the data and the corresponding quantum states:
\begin{equation}
    \mathbf{D}_{ij}^{'}=\frac{\arccos{(\langle\varphi_i\vert\varphi_j\rangle)}}{max(\mathbf{D}^{'})}, \quad
    {L}_{map}=\Sigma_{i=1}^n\Sigma_{j=1}^n\left\vert\mathbf{D}_{ij}-\mathbf{D}_{ij}^{'}\right\vert
    \label{mapping loss}
\end{equation}
where $\mathbf{D}_{ij}$ is the normalized Euclidean distance between samples $i$ and $j$. Similarly $\mathbf{D}_{ij}^{'}$ is the normalized distance defined in Eq.~\ref{mapping loss} between quantum state $i$ and $j$.

Compared with the existing mapping methods for  quantum algorithms, the main difference of the proposed mapping method is applying trainable parameters to maintain the distance relationships between data and quantum states, and  which reduces information loss. Our experiments will show the effectiveness of this  approach.

\subsection{Ego-graph Decompositional Processing Strategy}
\label{Decompositional Processing Strategy}
One bottleneck for quantum machine learning is the lack of available physical qubits in handling real-world data.

For efficient use of the available  qubits, we propose a decompositional processing strategy based on an  ego-graph decomposition. Concretely, according to Eq.~\ref{GIN} and Eq.~\ref{node representation}, the iterative updating of the representation of node $v$ is obtained by aggregating node $v$ together with its neighbors $\emph{N}(v)$ at the current iteration. As a result, the iterative updating of each node in a graph is only related to its neighbors. Therefore, we regard a node and its neighbors as an ego-graph. A graph containing $N$ nodes can be decomposed into $N$ ego-graph. The egoQGNN computes the representation of each ego-graph sequentially. As such, the process only requires a fixed number of qubits. 

To commence, the entire graph is divided into  ego-graph using a classical computer. The number of ego-graph is equal to the number of nodes. Next, the quantum circuits process each ego-graph sequentially and return their representations to the classical computer. Finally, the ego-graph representations are re-merged to reconstitute the original graph representation using the classical computer.

For a graph with $N$ nodes, the number of qubits required to compute its representation is accordingly reduced from $N$ to $D$ by introducing the above ego-graph decomposition, where $D$ is the maximum degree of the graph. If the number of available qubits $n<D$, we can divide the neighbour sets $\mathcal{N}(v)$ of node $v$ into $M$ sets: $\mathcal{N} (v)=\mathcal{N}_1(v)\cup\mathcal{N}_2(v)\cup .... \cup\mathcal{N}_m(v)$. All sets satisfy the conditions:
\begin{equation}
    \vert\mathcal{N}_i(v)\vert\le n, \quad 
    \mathcal{N}_i(v)\cap\mathcal{N}_j(v)=\varnothing, \quad i,j \in [1,m] , i\ne j
    \label{subsets}
\end{equation}

We sequentially compute the representations of the sets as follows: 
\begin{equation}
    \vert\varphi_{v{_i}}\rangle^{t-1}=\bigotimes_{\mu\in\mathcal{N}_i(v)}\mathbf{U}_2\vert\varphi_\mu\rangle^{t-1}
\end{equation}
\begin{equation}
    \vert\varphi_v\rangle^{t}=\mathbf{U}_1\vert\varphi_v\rangle^{t-1}\bigotimes\bigg(\bigotimes_{i\in[1,m]}\vert\varphi_{v{_i}}\rangle^{t-1}\bigg),\quad i \in [1,m] \label{mer_qs}
\end{equation}

Eq.~\ref{mer_qs} is computed using the classical computer, as $\vert\varphi_{v{_i}}\rangle^{t-1}$ is stored in the classical computer. 

\begin{algorithm}[tb!]
\caption{Decompositional processing for a graph}
\label{divide algo}
\begin{small}
\textbf{Input}: $G=(V,E)$, $\vert V\vert=N$; node features X=\{$x_v \vert \forall v \in \mathcal{V}$\}.\\
\textbf{Output}: ego-graph set $\{S_v \vert\forall v\in V\}$, ego-graph features set $\{C_v \vert\forall v\in V\}$.
\begin{algorithmic}[1]
 \FOR{$ v \in \mathcal{V}$}
        \STATE add $v$ and $N(v)$ to $S_{v}$;
        \FOR{$\mu\in N(v)$}
            \STATE add $X_{\mu}$ to $C_{v}$;
        \ENDFOR
    \ENDFOR
\end{algorithmic}
\end{small}
\end{algorithm}

For the effectiveness of the decompositional processing strategy, we provide the following results.

\begin{theorem}\label{theorem:2}
For the same inputs, Eq.~\ref{mer_qs} is equivalent to Eq.~\ref{node representation}.
\end{theorem}

We provide proof of this theorem in the Appendix.

\begin{algorithm}[tb!]
\caption{Trainable mapping for quantum representation}
\label{mapping algo}
\begin{small}
\textbf{Input}: input features X=\{$x_v\in\mathbb{R}^{d} \vert \forall v \in \mathcal{V}$\}, random initial parameters of mapping circuit P$\in \mathbb{R}^{d}$, the mapping circuit $MC$.\\
\textbf{Output}: trained initial parameters of mapping circuit $P_t\in \mathbb{R}^{d}$.
\begin{algorithmic}[1]
  \FOR{$ i\in V$}
        \FOR{$ j\in V$}
            \STATE $\mathbf{D}_{ij}=x_i\cdot x_j$;
        \ENDFOR
    \ENDFOR
    \FOR{$ i\in V$}
        \STATE $\vert\varphi_i\rangle\gets MC(P,x_i)$;
    \ENDFOR
    \FOR{$ i\in V$}
        \FOR{$ j\in V$}
            \STATE $\mathbf{D}_{ij}^{'}=\arccos{(\langle\varphi_i\vert\varphi_j\rangle)}$;\\
         $L_{map}=L_{map}+\vert\mathbf{D}_{ij}-\mathbf{D}_{ij}^{'}\vert$;
        \ENDFOR
    \ENDFOR
    $P_t=\mathop{\argmin}\limits_{P}  Loss$;
\end{algorithmic}
\end{small}
\end{algorithm}

\subsection{Structure of  the egoQGNN}
The output of the quantum circuit is a tensor product over the quantum states of all  the nodes, giving a high-dimensional vector. To classify graphs, their von Neumann entropy is summed over the quantum states of all nodes and used  as a characterization of the graph.
 \begin{figure*}[tb!]
    \centering
    \includegraphics[scale=0.27]{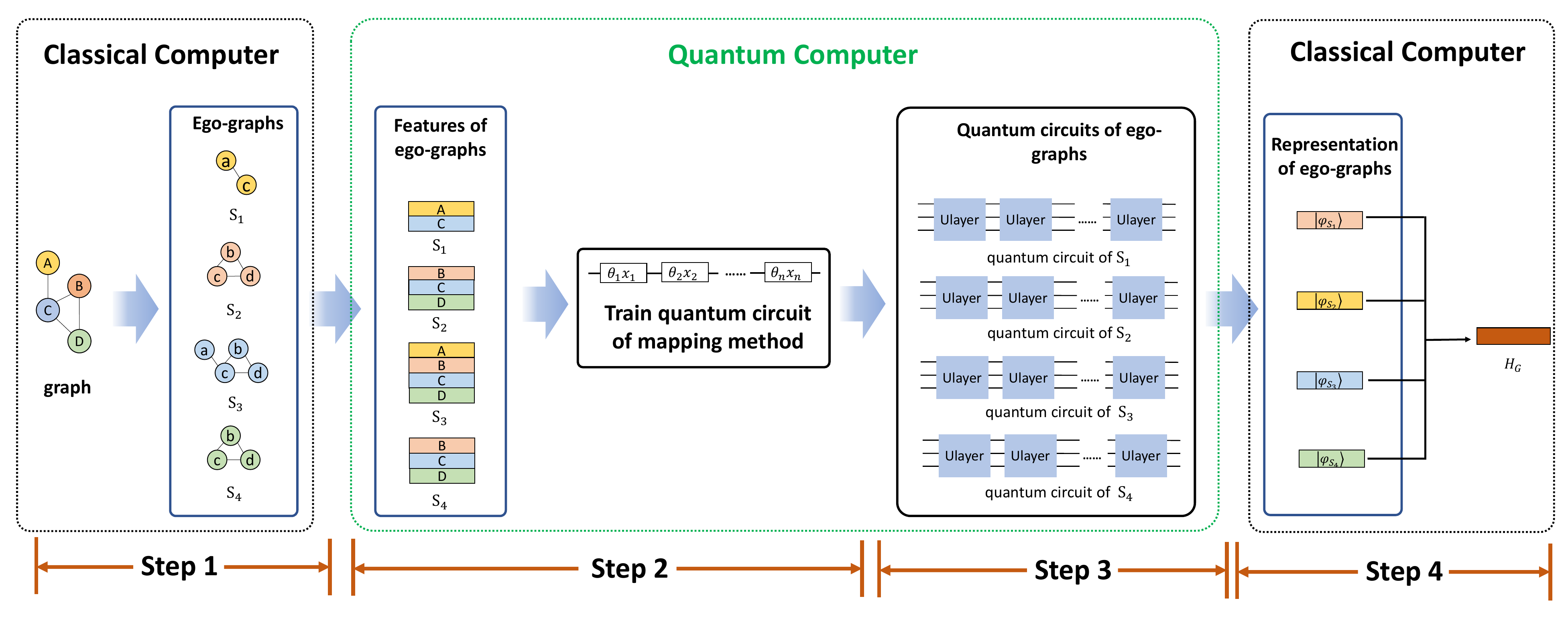}
    \vspace{-10pt}
    \caption{An instance of the egoQGNN. Step 1 and Step 4 are implemented on a classical computer. The quantum circuits of Step 2 and Step 3 can run on a quantum circuit or simulator of a classical computer.}
    \label{egoQGNN}
\vspace{-10pt}
\end{figure*}

The classical Shannon entropy measures the uncertainty associated with a classical probability distribution for a set of data. Quantum states can be described in a similar way using density operators in place of probability distributions, and the von Neumann entropy in place of the Shannon entropy. For a system with pure quantum state $\vert\varphi_i\rangle$ the density matrix
is $\mathbf{\rho}=\sum_i \vert \varphi_i\rangle \langle \varphi_i \vert$, and the graph representation $\mathbf{h}_G$ is:
\begin{equation}
    \mathbf{h}_G=\Sigma_{v\in G}S(\mathbf{\rho}_v),\quad \mathbf{\rho}_{v}=\vert\varphi_{v}\rangle\langle\varphi_{v}\vert
\end{equation}
\begin{equation}
    S(\mathbf{\rho}_v)=-tr(\mathbf{\rho}_v \log\mathbf{\rho}_v)
    \label{classification}
\end{equation}

For the binary classification problem, the Quantum Support Vector Machine (QSVM) \cite{Havlicek2019} classifies data by measuring components of quantum states along the Pauli-Z direction. The quantum states on the upper Bloch hemisphere are assigned to one class and the quantum states on the lower Bloch hemisphere are assigned to the other class. Our approach is essentially the same as QSVM. We use two quantum states corresponding to two classes, and these correspond to the upper and lower Bloch hemispheres. The von Neumann entropies of two quantum states are $\mathbf{\rho}_0$ and $\mathbf{\rho}_1$. Suppose $\mathbf{h}_G$ is the representation of graph $G$ to be classified, the label $y_G$ of graph $G$ is:
\begin{equation}
    y_G=
    \begin{cases}
    0, \quad if \mid\mathbf{\rho}_0-\mathbf{h}_G\mid\ge\mid\mathbf{\rho}_1-\mathbf{h}_G\mid\\
    1, \quad else
    \end{cases}
    \label{graph label}
\end{equation}

More details about the training of the egoQGNN are given in the Appendix. The egoQGNN consists of the following processing steps, where $K$ is the number of iterations.

\textbf{Step 1)} Classical  computer decomposes graph into ego-networks.

\textbf{Step 2)} Train the circuit of the mapping method with node features. The trained parameters of the circuit are stored on a classical computer. After training, the node features of the ego-networks are mapped into quantum states by applying the circuit.

\textbf{Step 3)} The quantum device runs the quantum circuit to compute the representations of the different ego-networks sequentially. The classical computer stores all the representations computed in this way.

\textbf{Step 4)} A classical computer computes the entropies of the individual representations and combines them using Eq.~\ref{classification}.

\textbf{Step 5)} \textbf{Steps 1-4} are repeated to obtain graph representations for classification. Fig.~\ref{egoQGNN} shows the structure.

\begin{algorithm}[tb!]
\caption{Framework of our egoQGNN}
\label{egoQGNN framework}
\begin{small}
\textbf{Input}: Ego-graph set $\{S_v \vert\forall v\in V\}$ with features $\{C_v \vert\forall v\in V\}$, the quantum circuit of egoQGNN mentioned in Sec.~\ref{quantum circuit} $QC$, Predefined von Neumann entropies of two classes: $\mathbf{\rho}_0$ and $\mathbf{\rho}_1$.\\
\textbf{Output}: label prediction for graph $G$: $l_G$.
\begin{algorithmic}[1]
    \FOR{$ v\in V$}
        \STATE $\vert\varphi_v\rangle\gets QC(S_v,C_v)$;\\
        $\mathbf{\rho}_v=\vert\varphi_v\rangle\langle\varphi_v\vert$;
    \ENDFOR\\
    \STATE$\mathbf{h}_G=\mathop{\Sigma}\limits_{v\in V}\mathbf{\rho}_v$;
   \STATE $y_G =
    \begin{cases}
    0, \quad if \mid\mathbf{\rho}_0-\mathbf{h}_G\mid\ge\mid\mathbf{\rho}_1-\mathbf{h}_G\mid;\\
    1, \quad else
    \end{cases}$
\end{algorithmic}
\end{small}
\end{algorithm}

\subsection{Summary of the egoQGNN Framework}
We summarize the framework of egoQGNN, whose decompositional processing strategy is shown in Alg.~\ref{divide algo}. This  corresponds to step 1 of Fig.~\ref{egoQGNN},  running on a classical computer. Alg.~\ref{mapping algo} involves step 2 of Fig.~\ref{egoQGNN}. After training, Alg.~\ref{mapping algo} outputs the initial parameters of the mapping circuit. Step 3 and step 4 of Fig.~\ref{egoQGNN} are summarized in Alg.~\ref{egoQGNN framework}, whose output is the label assigned to the  input graph.

\section{Experiments}
In this section, we perform evaluations of the proposed egoQGNN model on the graph classification task. We compare  the egoQGNN to both state-of-the-art graph kernels and deep learning methods. We conduct experiments on six standard graph classification benchmarks.
\begin{table*}[tb!]
\centering
\caption{\centering{Statistics of the used graph datasets.}}
\resizebox{0.85\textwidth}{!}{
\begin{tabular}{lcccccc}
\\\hline
Datasets&MUTAG&PTC\_MR&PTC\_FM&PTC\_MM&PTC\_FR&PROTEINS\\\hline
Max \# nodes& 28& 109& 64& 64& 64& 620\\\hline
Mean\#nodes&17.9&14.1&14.1&14.0&14.5&39.1\\\hline
Mean \# edges& 19.8& 14.7& 29.0& 28.63& 30.0& 72.82\\ \hline
\# graphs& 188& 344& 349& 336& 351& 1113\\ \hline
\# node labels& 6& 18& 18& 20& 19& 61\\ \hline
\# edge labels& 3& 4& 4& 4& 4& -\\ \hline
\# classes& 2& 2& 2& 2& 2& 2\\ \hline
\end{tabular}
}
\label{tab2}
\end{table*}

\begin{table*}[tb!]%\footnotesize
\centering
\caption{\centering{Evaluation for graph classification accuracy over six benchmarks (in \% $\pm$ standard error).}}
\vspace{-15pt} 
\resizebox{1.0\textwidth}{!}{
\begin{tabular}{rllllll}
\\\hline
Datasets &MUTAG &PTC\_FM &PTC\_FR &PTC\_MM &PTC\_MR &PROTEINS \\ \hline
WL~\cite{2010Weisfeiler}& \bm{$90.4\pm5.7$}& $60.4$& $65.7$& $66.6$& $59.9\pm4.3$& $75.0\pm3.1$\\
CSM~\cite{articleCSM}& $85.4$& $63.8$& $65.5$& $63.3$& $58.1$& -\\
DGCNN~\cite{zhang2018end}& $85.8$& $57.31\pm4.2$& $63.5\pm3.9$& $60.96\pm8.4$& $58.6$& $70.9\pm2.8$\\
R-GCN~\cite{2018ModelingR-GCN}& $81.5$& $60.7$& $65.8$& $64.7$& $58.2$& -\\
edGNN~\cite{xu2018powerful}& $86.9$& $59.8$& $65.7$& $64.4$& $56.3$& -\\
GIN~\cite{jaume2019edgnn}& $89.4\pm5.6$& $64.3\pm10.0$& $65.3\pm5.6$& $64.8\pm5.9$& $64.6\pm7.0$& $76.2\pm2.8$\\
RW-GNN~\cite{nikolentzos2020random}& $89.2\pm4.3$& $61.0\pm1.5$& $63.0\pm2.1$& $63.1\pm1.9$& $56.1\pm1.2$& $74.7\pm3.3$\\
DropGNN~\cite{DropGNN}& $90.4\pm7.0$& $63.2\pm5.1$& $66.3\pm8.6$& $64.8\pm4.5$& $65.1\pm5.2$& $76.3\pm6.1$\\
IEGN~\cite{IEGN}& $84.6\pm10.0$& $60.8\pm3.0$& $59.8\pm1.4$& $61.1\pm2.1$& $59.5\pm7.3$& $75.2\pm4.3$\\\hline
Gra+QSVM~\cite{Havlicek2019}& $81.80\pm10.3$& $60.66\pm5.4$& $62.50\pm10.2$& $57.50\pm8.8$& $60.89\pm7.3 $&$71.10\pm6.6$  \\
Gra+QCNN~\cite{Cong2018}& $81.04\pm14.1$& $59.86\pm10.9$& $63.2\pm11.2$& $58.96\pm9.9$& $61.53\pm8.6$&$77.05\pm10.1$  \\
Gra+QCNN (w/ M)~\cite{Cong2018}&$81.88\pm9.3$& $61.88\pm3.5$& $64.8\pm4.11$& $61.25\pm5.53$ 1&$62.02\pm3.3$& $78.9\pm7.8$ \\
GraphQNTK~\cite{GraphQNTK}& $88.4\pm6.5$& -& -& -& $62.9\pm5.0$& $71.1\pm3.2$ \\\hline
egoQGNN (w/o M)& $82.28\pm{10.53}$& $60.35\pm6.4$& $64.40\pm9.4$& $60.71\pm2.2$& $62.84\pm8.3$& $70.51\pm5.6$\\
egoQGNN& $85.83\pm3.7$ & \bm{$64.57\pm4.6$}& \bm{$67.3\pm5.6$}& \bm{$65.67\pm2.5$}& \bm{$65.14\pm5.6$}&\bm{$79.78\pm4.7$} \\ \hline
\end{tabular}}
\label{tab3}
\vspace{-10pt}
\end{table*}

\subsection{Datasets and Baselines}

An overview summary of the datasets used in our experiments  is given in Table \ref{tab3}. The data includes MUTAG \cite{1991Structure}, four variants of PTC \cite{2001The} and PROTEINS \cite{R1990Protein}. For all of the datasets, the nodes have the  categorical input features  required for egoQGNN.

\begin{itemize}
    \item[$\bullet$] MUTAG: The graphs contained in  this dataset represent  heteroaromatic nitro or mutagenic aromatic compounds. The nodes and edges represent atoms and chemical bonds, respectively. The nodes labels represent the chemical identity of the atoms. All of the graphs in this dataset belong to one of two classes which  represent whether the graph has a mutagenic effect or not.
    \item[$\bullet$] PTC: The Predictive Toxicology Challenge (PTC) has four variants, representing molecule carcinogenicity on male mice (PTC\_MM), male rats (PTC\_MR), female mice (PTC\_FM), female rats (PTC\_FR), respectively. The graphs from each variant are labeled by their carcinogenicity on male and female mice and rats. 
    \item[$\bullet$] PROTEINS: This dataset includes two classes of proteins. The first is enzymes, while the second is non-enzymes. Nodes represent the different amino acids belonging to the proteins. Edges  represent whether  the distance between pairs of  amino acids is less than 0.6 nanometres.
\end{itemize}

To the best of our knowledge, GraphQNTK~\cite{GraphQNTK} is one of the few currently available (and also the most up-to-date) quantum algorithms that can be applied to  realistically sized real-world graph datasets. In fact, for the  typical quantum algorithms, QSVM \cite{Havlicek2019} and QCNN \cite{2018Modeling}, due to the limitations on the number of qubits, high-dimensional graph data cannot be handled directly. Fortunately, Yanardag et al. \cite{yanardag2015deep} proposed a  method based on the number of graphlets to encode high-dimensional graphs using  low-dimensional representations. We, therefore, use this method to encode a high-dimensional graph as  an 8-dimensional vector and then use this vector as the input  to both QSVM and QCNN. Additionally, we compare  egoQGNN with several state-of-the-art baselines for graph classification: 

(1) The kernel based models: WL subtree kernel  \cite{2010Weisfeiler} and Subgraph Matching Kernel (CSM) \cite{articleCSM}.

(2) The state-of-the-art GNNs: These  include Diffusion convolutional neural networks (DCNN) \cite{2015Diffusion}, Deep Graph CNN (DGCNN) \cite{zhang2018end} and Relational Graph Convolutional Networks (R-GCN) \cite{2018ModelingR-GCN}, Graph Isomorphism Network (GIN) \cite{xu2018powerful} and  edGNN \cite{jaume2019edgnn}, Random Walk Graph Neural Networks (RW-GNN)\cite{nikolentzos2020random}, Dropout Graph Neural Networks(DropGNN)\cite{DropGNN}, Invariant-Equivariant Graph Network(IEGN)\cite{IEGN}.

%Experimental Setup
\subsection{Experimental Setup}
For experiments, we use  three Ulayers in egoQGNN. All three Ulayers have identical structures but no shared parameters. In the quantum circuit of the mapping method, the feature of a node can be mapped to several qubits. In experiments, we map the  feature  of each node into a qubit since the node features of the datasets are simple (a scalar).
%Erh somethin missing above
The quantum circuits of the proposed method are similar to QCNN and QSVM,  and RX, RY, RZ gates are applied on each qubit sequentially.
During the training, UOBYQA  \cite{2002UOBYQA}, which is based on a derivative-free optimization method, is used to optimize the egoQGNN. 
%Mcclean et al.  \cite{Mcclean} have  drawn attention to the gradient varnishing problem for gradient descent methods when used for quantum machine learning. Besides, since we use a  decompositional processing strategy in  DQGNN, providing an explicit expression for the gradient is difficult. 

Due to the lack of qubits, QSVM only accepts 2-dimensional or 3-dimensional data. So, we use PCA to transform an 8-dimensional graphlet 
%ERH: what do you mean by number? Subgraph frequency counts?
frequency counts vector into a 3-dimensional vector. The quantum machine learning methods  are run in  simulation on a  classical computer. The code for QSVM is provided by Qiskit \cite{Qiskit}, and  the code for  QCNN is provided by Tensorflow Quantum \cite{TFQ}. 

The results for the  deep learning methods mostly come from the existing studies \cite{jaume2019edgnn} \cite{xu2018powerful}. Using  the available codes provided by the authors, we perform 10-fold cross-validation to compute  the accuracies of GIN and DGCNN on the PTC\_FM, PTC\_MM and  PTC\_FR datasets, and both  R-GCN and edGNN on  the PROTEINS dataset. The parameters for the  deep learning methods are those  provided by the authors. 

For fairness, all of the  methods compared  are run on the same computing  device, namely  an  Intel Xeon CPU E3-1270 v5 with 32GB RAM. The proposed method is implemented using  the OriginQ\cite{OriginQ} simulator on a classical computer. As yet existing quantum coding platforms\cite{Qiskit,TFQ}  are unable to provide the  fast and effective interaction between a  classical computer and  a quantum computer  required by the proposed framework. For this reason,  all  of the reported experiments are performed using a  simulator running on a classical computer.

\subsection{Results and Discussion}
The performances on graph classification are assessed in terms of accuracy. We report the average and standard deviation of the 10-fold  validation accuracies. The results achieved by egoQGNN are reported in Table \ref{tab3}.
We also give the performance achieved by egoQGNN without the trainable mapping method. This is done  to demonstrate the effectiveness of the mapping strategy described  in Section \uppercase\expandafter{\romannumeral4}.

% Compare with machine learning methods
\subsubsection{Comparison with machine learning methods}
For the evaluation, we employ the same structure for the proposed egoQGNN model on all of the graph datasets studied. Results in Table \ref{tab3} indicate that egoQGNN achieves the best results on 5 out of the 7 benchmarks,  showing  that in many cases a clear improvement is obtained with respect to the GNN models. The accuracies of the egoQGNN with a trainable mapping method on PTC\_FM and PTC\_MM are $64.57\pm4.64$ and  $65.67\pm2.54$ respectively, which are roughly equivalent  to less powerful methods. For PTC\_FR and PTC\_MR, egoQGNN achieves $1.96$ and $0.54$ improvements over the next-best  performing methods. Besides, even in the cases where egoQGNN does not achieve top performance, its accuracy is close to that of the  GNNs studied. 

% Compare with other quantum machine learning methods
\subsubsection{Comparison with alternative quantum machine learning methods} 
Gra+QSVM and Gra+QCNN in Tabel \ref{tab3} refer to using the graphlet count vector as the input to QSVM and QCNN, respectively. QSVM has its mapping circuit shown in  the Appendix but QCNN does not. We apply the mapping circuits shown in Fig.~\ref{Ulayer} to QCNN (Gra+QCNN (w/ M)). The results of Gra+QSVM, Gra+QCNN and Gra+QCNN (w/ M) represent no improvement on egoQGNN. Moreover, compared to Gra+QCNN, Gra+QCNN (w/ M) achieves higher accuracy. Notably, compared to GraphQNTK, our model performs better on all datasets except MUTAG. The reason for this may be the hierarchical  structure of our model, which is not adopted in GraphQNTK. This demonstrates that our proposed trainable mapping method is also effective when combined with alternative quantum machine learning algorithms.

\subsubsection{Comparison of egoQGNN (w/o M) and egoQGNN}
We observe that egoQGNN  when combined with our trainable mapping method slightly and consistently outperforms egoQGNN without this trainable mapping i.e. egoQGNN (w/o M). Since they have the same structure, the improvement may be attributed to less information loss compared to the egoQGNN without the mapping method. Note that the accuracy of egoQGNN is about 9\% higher than egoQGNN (w/o M) on the PROTEINS dataset, while for the MUTAG dataset, the accuracy of egoQGNN (w/o M) is close to that of egoQGNN. Table \ref{tab3} shows that PROTEINS uses 61 node labels while MUTAG uses only seven. This means that the  PROTEINS dataset  requires higher dimensionality  for the  node features and suffers more serious information loss. The results show that our trainable mapping method reduces information loss and improves performance. Besides, for most datasets, the standard error for egoQGNN is less than that for  egoQGNN (w/o M). For example, the standard errors for  egoQGNN (w/o M) and egoQGNN on PTC\_FR are respectively $9.4$ and $5.6$. When there is no trainable mapping, the elements of the node feature vector are used as the gate parameters to map data to quantum states. This leads to a random distribution of quantum states in the Hilbert space.

% Comparison of parameters
\subsubsection{Comparison of parameters}
Compared with the deep learning methods, egoQGNN has fewer parameters but achieves comparable performance, as shown in Table \ref{tab4}. One of the possible reasons is that as mentioned in \cite{Maria2019}, the Hilbert space is a high-dimensional space, and the performance of quantum machine learning algorithms can be improved even though their parameters are fewer in number. For example, for a 32-dimension input, a layer of a GNN model requires a $32\times32$ weight matrix to transform the input to  a 32-dimension output. As a result, this layer has 1024 parameters. For quantum machine learning, the entanglement of 5 qubits generates  32-dimensional quantum states. Suppose that three quantum gates are applied to each qubit and each quantum gate has a parameter.  In this instance, the quantum machine learning model transforms a 32-dimensional quantum state to a new 32-dimensional quantum state as output using  only 15 parameters. Besides, similar to \cite{chami2019hyperbolic}, egoQGNN captures the feature of the  nodes in a  non-Euclidean space. This reduces the distortion and leads to an improvement in performance. 

\subsubsection{Comparison of run-times}
\begin{figure}[t!]
    \centering
    \includegraphics[scale=0.35]{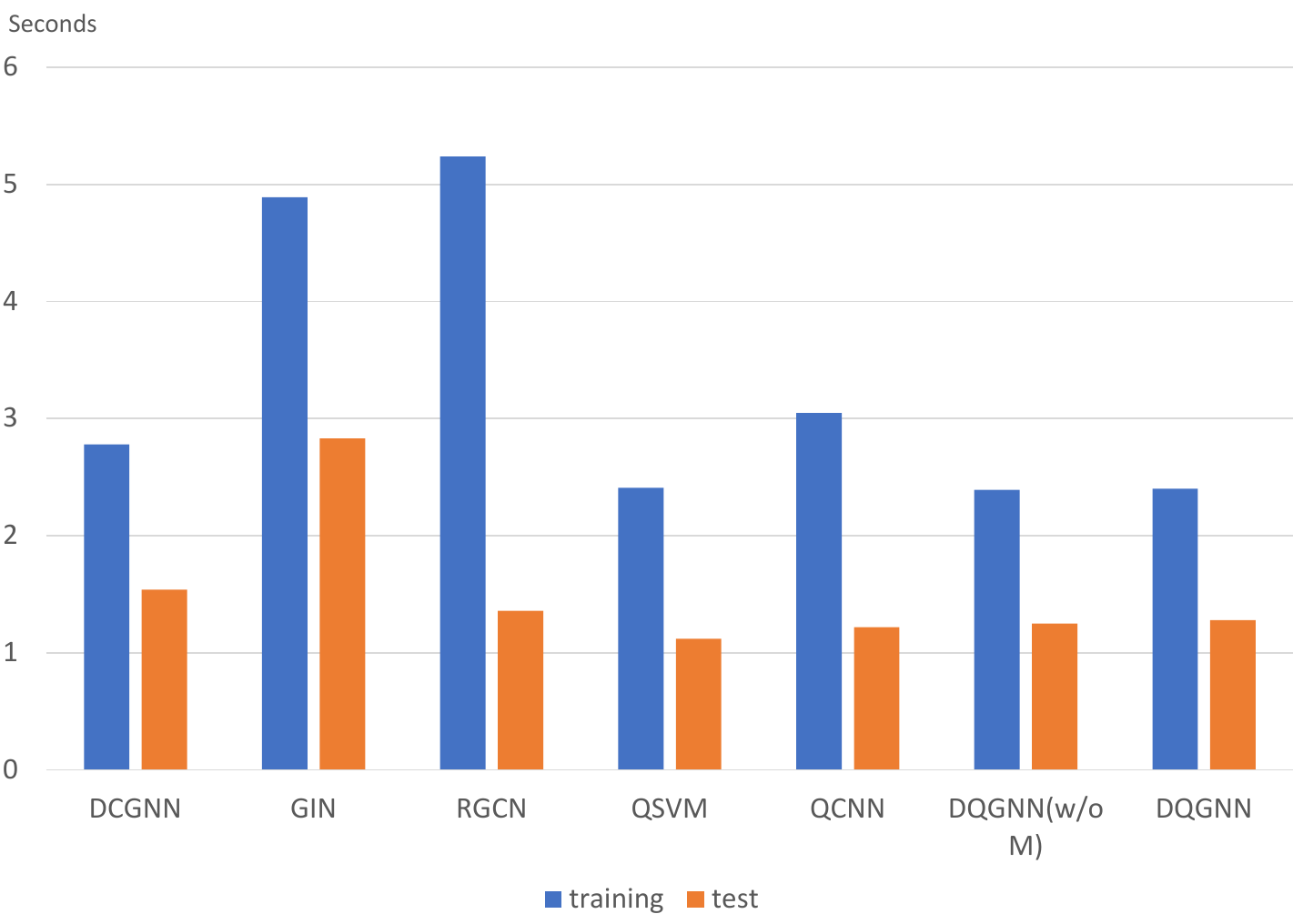}
     \caption{Running time comparison on MUTAG. The blue and orange bars indicate the per-epoch training time and test time respectively.}
    \label{Fig12}
\end{figure}

We report the average run times of egoQGNN and the baselines on MUTAG in Fig \ref{Fig12}. For fairness, all  the methods run on a machine with an Intel Xeon CPU E3-1270 v5. The run-time of the proposed method is comparable to the alternative   quantum methods and superior to several of the  deep learning methods. One important observation is that egoQGNN is less time-consuming than QCNN and close in performance  to QSVM. It is worth noting that deep learning based methods, i.e. DCGNN, GIN and RGCN are no faster than  the quantum computing methods since the comparison is executed on a conventional  CPU-based  machine.

\begin{table}[tb!]
\centering
\caption{\centering{Quantum model parameter size (the number of parameters) comparison. The parameters of GNN include all the parameters of the weight matrix to be trained. The parameters of the  quantum algorithm are the angles required by all of the quantum gates (unitary matrix).}}
\vspace{-15pt}
\resizebox{0.35\textwidth}{!}{
\begin{tabular}{lr}
\\\hline
Quantum model& Parameter size\\\hline
DGCNN~\cite{zhang2018end}& 2,560\\ 
R-GCN~\cite{2018ModelingR-GCN}& 16,704\\ 
edGNN~\cite{xu2018powerful}& 9,345\\ 
GIN~\cite{jaume2019edgnn}& 13,000\\\hline
QSVM~\cite{Havlicek2019}& 36\\
QCNN~\cite{Cong2018}& 54\\\hline
egoQGNN (w/o M)& 36\\
egoQGNN& 43\\\hline
\end{tabular}
}
\label{tab4}
\end{table}

\section{Conclusions and Outlook}
In this paper, we have  developed a novel hybrid quantum-classical algorithm for graph-structured data, namely  the Ego-graph based Quantum Graph Neural Network (egoQGNN). We have  introduced the theoretical framework of the  egoQGNN and provided mathematical proof concerning its ability to identify graph isomorphisms. We also propose a decompositional processing strategy, which liberates egoQGNN from the limitation of the number of  qubits. With the aid of classical computers, egoQGNN can handle graphs with larger sizes as input on a quantum device of a given size. Moreover, to reduce information loss during  the mapping of data to quantum states, a trainable method is proposed. Experimental results demonstrate this method is  beneficial  and leads to improvements in  the performance of quantum machine learning algorithms. There are several potentially interesting directions for future work. As a long-term goal, quantum techniques need to be utilized to achieve exponential speed-up.In  the short term, given the  fast methodological developments~\cite{ZhangTNNLS21,ZhangTNNLS22,WuICLR22} on the  node-level embedding with classic computers, it is also imperative to develop competitive quantum counterparts that are feasible on near-term quantum devices. A natural extension for quantum node embedding is to develop quantum  solvers for the problem of combinatorial optimization e.g. graph matching~\cite{WangTPAMI21,WangTPAMI22}, since  there are a few quantum solvers~\cite{birdal2021cvpr} and these  are yet not learnable.

%{\color{red}{There are two main limitations of the proposed method. Firstly, egoQGNN is incapable of handling node classification since mapping a $n$-dimensional node feature into quantum state of a single qubit may lead to great information loss. The whole graph may robustness to such information loss while a single node may not. Secondly, egoQGNN requires time-consuming transmission of quantum computer and traditional computer. For instance, the computation of Eq.~\eqref{next representation} needs to transfer snapshots of quantum states from quantum computer to classical computer. If the quantum circuits could complete the computation of Eq.~\eqref{next representation} in the future, this limitation would be alleviated.}}

 \section*{Acknowledgment}
 This work was partly supported by National Key Research and Development Program of China (2020AAA0107600), National Natural Science Foundation of China (62176227, U2066213 and 62222607), the fundamental research funds for the central universities under grant 2072021004 and Shanghai Municipal Science and Technology Project (22511105100).

\appendix

\section{Proof and Analysis}
\textbf{Proof of the Lemma 1:} 

For four different quantum states $A, B, C, D$ satisfying the following condition:
\begin{equation}
    A\otimes B=C\otimes D
\end{equation}

Suppose $A, C$ are $n-$ dimensional quantum states, $B, D$ are $m-$ dimensional quantum states. The $a_i, b_j, c_i, d_j$ are $i$-th or $j$-th components of $A, B, C, D$ respectively. So, the above formula can be rewritten as:
\begin{equation}
    a_ib_j=c_id_j \quad (i,j\in N^*, i\le n, j\le m)
\end{equation}

So, for any $k\in[1,m], k\in N^*$:
\begin{equation}
    \begin{cases}
    a_1b_k=c_1d_k\\
    a_2b_k=c_2d_k\\
    \quad\quad\vdots\\
    a_nb_k=c_nd_k
    \end{cases}\label{mul_com}
\end{equation}

Obviously, the $A, C$ are linear correlations. Suppose $d_k/b_k=\varepsilon$, $\varepsilon\in\mathbb{R}$:
\begin{gather}
    a_i/c_i=d_k/b_k=\varepsilon\\
    a_i=\varepsilon\cdot c_i\\
    A=\varepsilon\cdot C
\end{gather}

Thus, the $B, D$ are also linear correlations.
\begin{gather}
    A\otimes B=C\otimes D\\
    \varepsilon\cdot C\otimes B=C\otimes D
\end{gather}

Eq.\eqref{mul_com} can be rewritten as:
\begin{equation}
    \begin{cases}
    \varepsilon\cdot c_1b_k=c_1d_k\\
    \varepsilon\cdot c_2b_k=c_2d_k\\
    \quad\quad\vdots\\
    \varepsilon\cdot c_nb_k=c_nd_k
    \end{cases}
\end{equation}

Naturally, $B=\frac{1}{\varepsilon}\cdot D$, the $B, D$ are linear correlation. This means that the tensor product does not  map non-zero vectors to the same representation unless they are linearly correlated.

\textbf{Proof of the Lemma 2:} 

For two linearly dependent quantum states: $A$ and $B$, $B=k\cdot A, k\in\mathbb{R}$. Suppose $A$ and $B$ are $n-$dimensional quantum states, the $a_i$ and $b_i$ are the $i$-th components of $A$ and $B$ respectively. According to the property of quantum states: 
\begin{equation}
    \Sigma_{i=1}^{n}\vert a_i\vert^2=\Sigma_{i=1}^{n}\vert b_i\vert^2=1
\end{equation}

$A$ and $B$ are linearly dependent quantum states and $B=k\cdot A$. Thus, the above formula can be rewritten as:
\begin{equation}
    \Sigma_{i=1}^{n}\vert b_i\vert^2=\Sigma_{i=1}^{n}\vert k \cdot a_i\vert^2=k^2 \Sigma_{i=1}^{n}\vert a_i\vert^2=\Sigma_{i=1}^{n}\vert a_i\vert^2
\end{equation}

So, $k=1$ or $k=-1$. It means that if $A$ and $B$ are linearly dependent quantum states, $\vert A\vert=\vert B\vert$. 

\textbf{Proof of the Lemma 3:}

For nodes $v$ and $\mu$, suppose $\vert\varphi_v\rangle^{t}$ and $\vert\varphi_\mu\rangle^{t}$ are quantum states of $v$ and $\mu$ in $t$-th layer respectively. There are two situations for $v$ and $\mu$:

\begin{itemize}
    \item [(1)] $v$ and $\mu$ have different numbers of neighbors. 
    
    If $v$ and $\mu$ have different numbers of neighbors, the dimensions of $\vert\varphi_v\rangle$ and $\vert\varphi_\mu\rangle^{t-1}$ are different. The $\vert\varphi_v\rangle^{t-1}$ and $\vert\varphi_\mu\rangle^{t-1}$ are 2-dimensional quantum states, because each node is represented by a qubit. The quantum state of a qubit is 2-dimensional. So, if $v$ and $\mu$ have different numbers of neighbors, the dimensions of $\vert\varphi_v\rangle^{t}$ and $\vert\varphi_\mu\rangle^{t}$ are different,  $\vert\varphi_v\rangle^{t}\neq\vert\varphi_\mu\rangle^{t}$. 
    
    \item [(2)] $v$ and $\mu$ have the same number of neighbors.
    
    The Eq. 10 can be rewritten as below:
    \begin{gather}
        \vert\varphi_v\rangle^{t}=\sigma(U_1\vert\varphi_v\rangle^{t-1}\bigotimes\vert\varphi_{\mathcal{N}_v}\rangle^{t-1})\\ 
        \vert\varphi_\mu\rangle^{t}=\sigma(U_1\vert\varphi_\mu\rangle^{t-1}\bigotimes\vert\varphi_{\mathcal{N}_\mu}\rangle^{t-1})\\
        \vert\varphi_{\mathcal{N}_v}\rangle^{t-1}=\bigotimes_{i\in\mathcal{N}(v)}U_2\vert\varphi_i\rangle^{t-1}\\
        \vert\varphi_{\mathcal{N}_\mu}\rangle^{t-1}=\bigotimes_{i\in\mathcal{N}(\mu)}U_2\vert\varphi_i\rangle^{t-1}\\
    \end{gather}
    
    According to Lemma 1 and Lemma 2, the tensor product maps two quantum states, $A$ and $B$, to the same representation if and only if $A=B$ or $A=-B$. Therefore,  $\vert\varphi_v\rangle^{t-1}\neq\vert\varphi_\mu\rangle^{t-1}$,  $\vert\varphi_v\rangle^{t}\neq\vert\varphi_\mu\rangle^{t}$. 
\end{itemize}

\textbf{Proof of Theorem 4:}

Suppose for two non-isomorphic graphs $G_1$ and $G_2$, the collections of representations of all nodes in the last layer of GNN are $\{h_v^K \vert v\in G_1\}$ and $\{h_\mu^K \vert \mu\in G_2\}$ respectively. Similarly, the collections of representations of all nodes in the last layer of egoQGNN are $\{ \vert\varphi_v\rangle^K \vert v\in G_1\}$ and $\{ \vert\varphi_\mu\rangle^K \vert \mu\in G_2\}$.  
If GNNs decide $G_1$ and $G_2$ are non-isomorphic, $\exists h_v^K \neq h_\mu^K\quad (v\in G_1, \mu\in G_2)$. According to Lemma 1-3,  for egoQGNN, $\exists \vert\varphi_v\rangle^K \neq \vert\varphi_\mu\rangle^K (v\in G_1, \mu\in G_2)$. Moreover, according to the Eq. 11:
\begin{gather}
    h_{G_1}=\Sigma_{v\in G_1}S(\rho_v), h_{G_2}=\Sigma_{\mu\in G_2}S(\rho_\mu)\\
    h_{G_1}\neq h_{G_2}
\end{gather}

The representations of $G_1$ and $G_2$ are unequal to each other. So, egoQGNN can distinguish graphs that are decided as non-isomorphic by GNNs.

\ifCLASSOPTIONcaptionsoff
  \newpage
\fi

\bibliographystyle{IEEEtran}
\bibliography{main}

\begin{IEEEbiography}[{\includegraphics[width=1in,height=1.25in,clip,keepaspectratio]{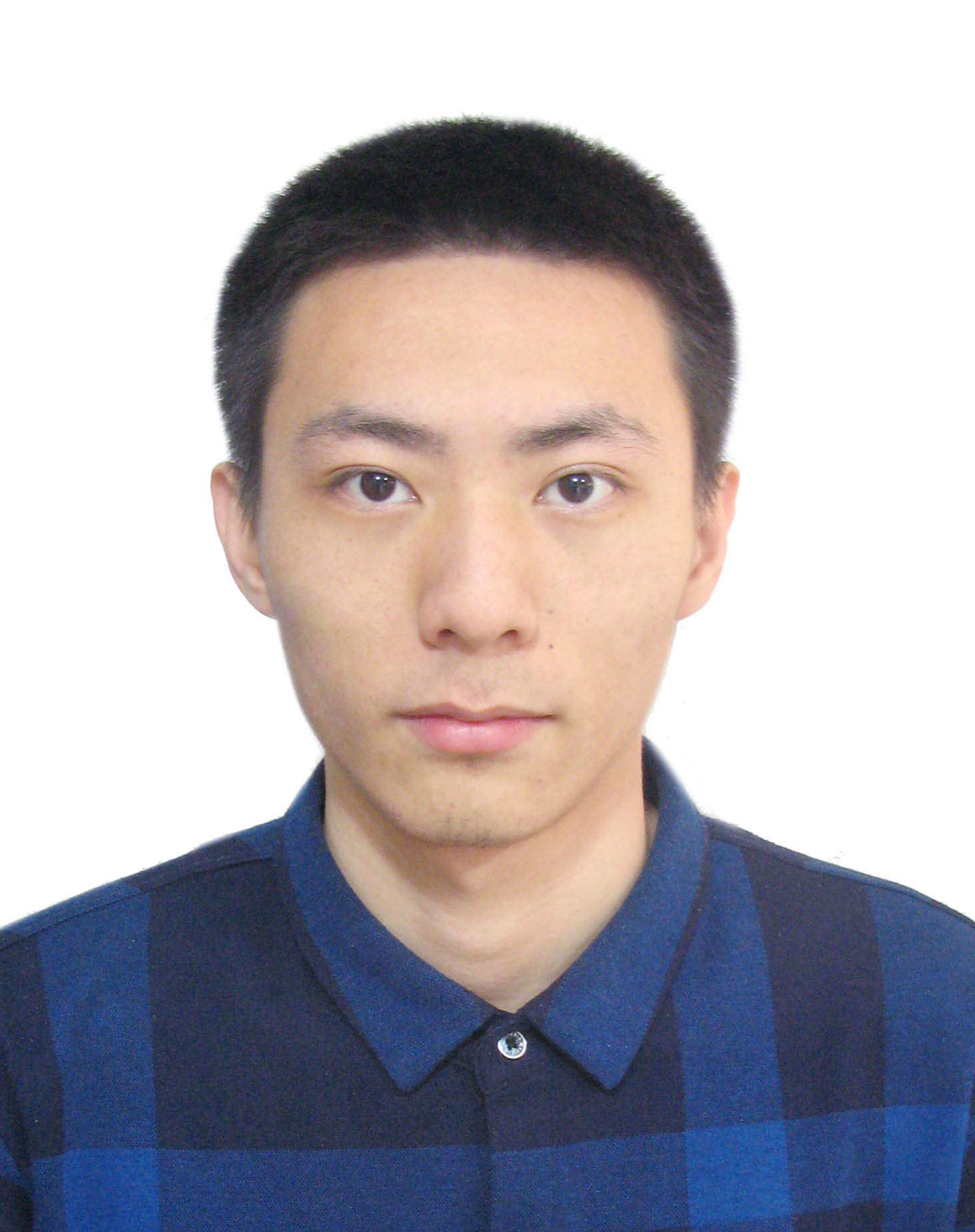}}]{Xing Ai} is now a PhD student at the Computing Department of Hong Kong Polytechnic University. He obtained his bachelor's degree from the Software Department of Xiamen University, China in 2015. He then received his master's degree from the Information Department of Xiamen University, China in 2019. His research interests include machine learning, graph representation, and quantum computing. He completed this paper during his study for his master's degree at Xiamen University. 
\end{IEEEbiography}

\begin{IEEEbiography}[{\includegraphics[width=1in,height=1.25in,clip,keepaspectratio]{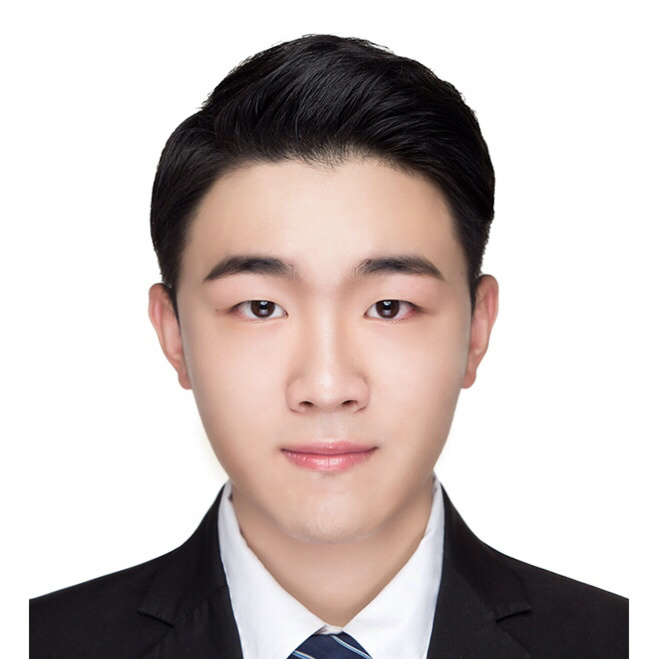}}]{Luzhe Sun}
is now a postgraduate student in the Department of Computer Science at the University of Chicago. His research interests include Graph Theory, Transform Learning, Quantum Computing. His recent work has focused on directed acyclic graph learning and secure trajectory planning and shared autonomy using causal inference. He completed this paper during his study for his bachelor's degree at Xiamen University. 

\end{IEEEbiography}

\begin{IEEEbiography}[{\includegraphics[width=1in,height=1.25in,clip,keepaspectratio]{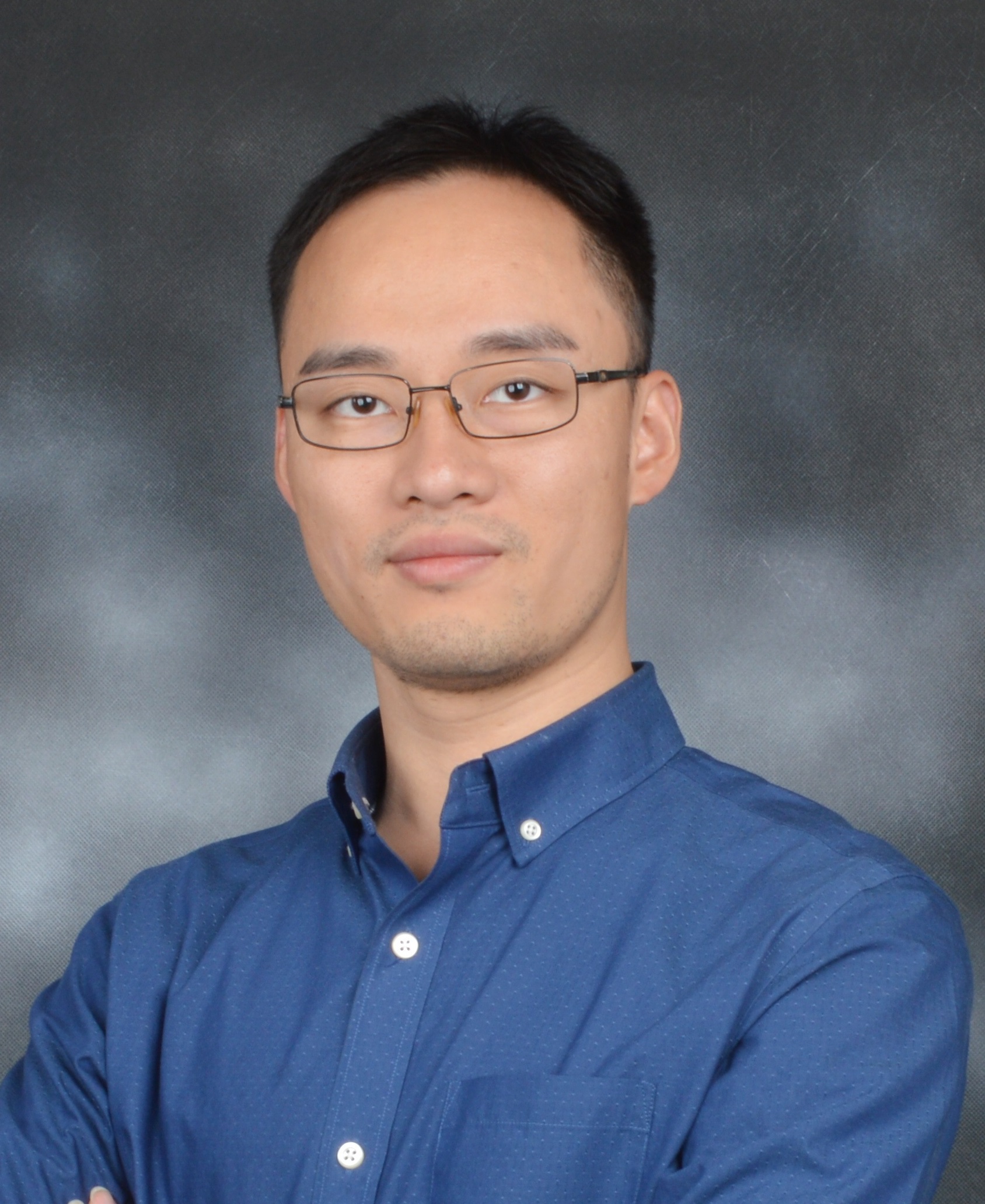}}]{Junchi Yan} (S'10-M'11-SM'21) is currently an Associate Professor with Department of Computer Science and Engineering, Shanghai Jiao Tong University, Shanghai, China. Before that, he was a Senior Research Staff Member and Principal Scientist with IBM Research where he started his career since April 2011. He obtained the Ph.D. at the Department of Electronic Engineering of Shanghai Jiao Tong University. His research interests are machine learning, as well as the intersection with combinatorial optimization and quantum computing. He has also recently published quantum graph learning works in SIGKDD and NeurIPS. He serves as Area Chair for NeurIPS/ICML/AAAI/CVPR, etc. and Associate Editor for Pattern Recognition.
\end{IEEEbiography}

\begin{IEEEbiography}[{\includegraphics[width=1in,height=1.25in,clip,keepaspectratio]{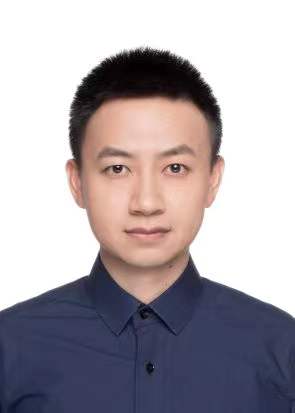}}]{Zhihong Zhang} received his BSc degree (1st class Hons.) in computer science from the University of Ulster, UK, in 2009 and the PhD degree in computer science from the University of York, UK, in 2013. He won the K. M. Stott prize for best thesis from the University of York in 2013. He is now an associate professor at the School of Informatics Xiamen University, China. His research interests are wide-reaching but mainly involve the areas of pattern recognition and machine learning, particularly problems involving graphs and networks. He is a recipient of the Best Paper Awards of the International Conference on Pattern Recognition ICPR 2018. He is currently an Associate Editor of Pattern Recognition Journal.
\end{IEEEbiography}

\begin{IEEEbiography}[{\includegraphics[width=1in,height=1.25in,clip,keepaspectratio]{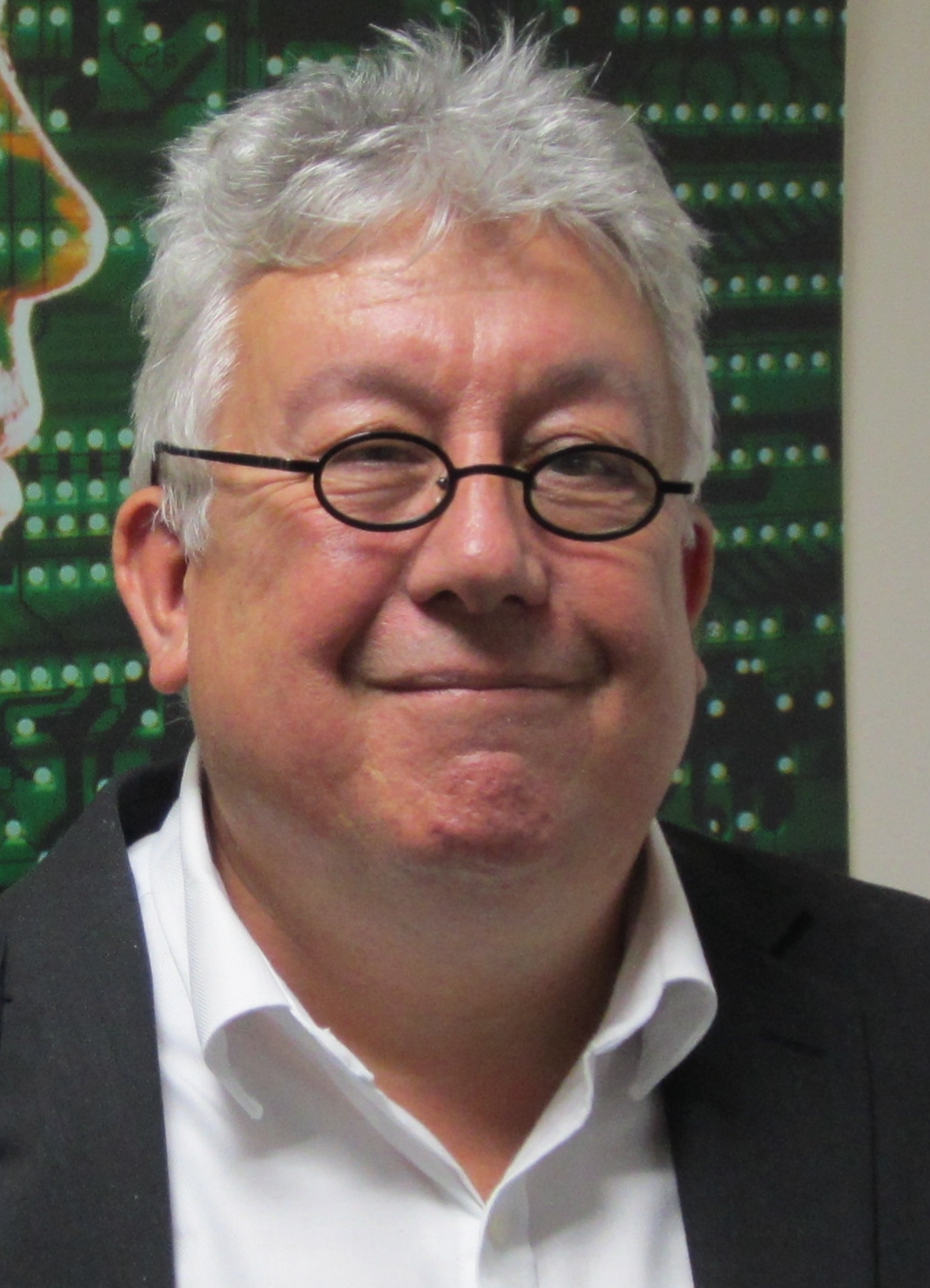}}]{Edwin R. Hancock} holds a B.Sc. degree in physics (1977), a Ph.D. degree in high-energy physics (1981) and a D.Sc. degree (2008) from the University of Durham, and a doctorate Honoris Causa from the University of Alicante in 2015. He is an Emeritus Professor in the Department of Computer Science at the University of York, Adjunct Professor at Beihang University and Distinguished Visiting Professor at Xiamen University.  His main research interests are in pattern recognition, machine learning and computer vision, where he has made sustained contributions to the use of graph-based methods and physics-based vision over the past 30 years. He was elected a Fellow of the Royal Academy of Engineering (the UK's national academy of engineering), in 2021. He is also a Fellow of both the International Association for Pattern Recognition and the IEEE. He was the 2016 Distinguished Fellow of the BMVA. He is currently Editor-in-Chief of the journal Pattern Recognition and was founding Editor-in-Chief of IET Computer Vision from 2006 until 2012. He has also been a member of the editorial boards of the journals IEEE Transactions on Pattern Analysis and Machine Intelligence, Pattern Recognition, Computer Vision and Image Understanding, Image and Vision Computing, and the International Journal of Complex Networks. He was Vice President of the IAPR from 2016 to 2018. He has been the recipient of the Pattern Recognition Medal (1992),  the IAPR Piero Zamperoni Award (2006), a Royal Society Wolfson  Research Merit Award (2008), and the IAPR  Pierre Devijver Award (2018), He is an IEEE Computer Society Distinguished Visitor for the period 2021-2023.
\end{IEEEbiography}


%\vfill


\end{document}